\documentclass[prd,preprint,superscriptaddress,preprintnumbers,eqsecnum,showpacs,nofootinbib,
nobibnotes,noeprint]{revtex4-1}
\usepackage[linkcolor=blue,citecolor=blue,urlcolor=blue,colorlinks=true,breaklinks]{hyperref} 

\usepackage{amsfonts,amssymb,amsmath,dsfont,epsfig}
\usepackage{amsfonts,amssymb,amsmath}
\usepackage{graphicx}
\usepackage{paralist}
\usepackage[english]{babel}
\usepackage{slashed}
\usepackage[usenames]{xcolor}
\usepackage{xspace}
\usepackage{float}
\usepackage{mathtools}
\usepackage{picture}
\usepackage[export]{adjustbox}

\usepackage{etoolbox}

\makeatletter
\patchcmd{\frontmatter@abstract@produce}
  {\vskip200\p@\@plus1fil
   \penalty-200\relax
   \vskip-200\p@\@plus-1fil}
  {}
  {}
  {}
\makeatother

\newcommand{\be}{\begin{equation}}
\newcommand{\bea}{\begin{eqnarray}}
\newcommand{\ee}{\end{equation}}
\newcommand{\eea}{\end{eqnarray}}
\newcommand{\euc}{{\s {\rm E}}}
\newcommand{\wic}{{\s {\rm E}}}
\newcommand{\sg}{{\!{sg}}}

\def\tl{0}

\newcommand{\Ls}{ \mathit{L}_{{sg}}}

\def\s#1{{\scriptscriptstyle #1}}

\usepackage[skins]{tcolorbox}
\usepackage{varwidth}

\makeatletter
\newtcolorbox{mybox}[1][]{
  enhanced,
  colframe=black, colback=white,
  sharp corners,
  boxrule=0.6pt,
}
\makeatother

\def\1eq#1{Eq.~(\ref{#1})}

\def\2eqs#1#2{Eqs.~(\ref{#1}) and~(\ref{#2})}
\def\3eqs#1#2#3{Eqs.~(\ref{#1}),~(\ref{#2}) and~(\ref{#3})}

\def\fig#1{Fig.~\ref{#1}}

\def\ie{{\it i.e.}, }
\def\eg{{\it e.g.}, }

\def\s#1{{\scriptscriptstyle #1}}

\def\MOMt{$\widetilde{\text{MOM}}$}

\def\g{\Gamma}

\newcommand{\fatg}{{\rm{I}}\!\Gamma}

\def\s#1{{\scriptscriptstyle #1}}

\begin{document}

\title{Infrared properties of the quark-gluon vertex \\ in general kinematics}

\author{A.~C.~Aguilar}
\affiliation{\mbox{University of Campinas - UNICAMP, Institute of Physics Gleb Wataghin,} 
13083-859 Campinas, S\~{a}o Paulo, Brazil.}

\author{M.~N. Ferreira}
\affiliation{\mbox{School of Physics, Nanjing University, Nanjing, Jiangsu 210093, China
}}
\affiliation{{Institute for Nonperturbative Physics, Nanjing University, Nanjing, Jiangsu 210093, China
}}

\author{G.T. Linhares}
\affiliation{\mbox{University of Campinas - UNICAMP, Institute of Physics Gleb Wataghin,} 13083-859 Campinas, S\~{a}o Paulo, Brazil.}

\author{B.~M. Oliveira}
\affiliation{\mbox{University of Campinas - UNICAMP, Institute of Physics Gleb Wataghin,} 13083-859 Campinas, S\~{a}o Paulo, Brazil.}

\author{J.~Papavassiliou}
\affiliation{\mbox{Department of Theoretical Physics and IFIC,} \\ University of Valencia and CSIC, E-46100, Valencia, Spain.}

\affiliation{ExtreMe Matter Institute EMMI,
GSI,  
Planckstrasse 1,
64291 Darmstadt,
Germany}

\begin{abstract}
In the present work 
we determine the eight form factors of the 
transversely-projected quark-gluon vertex in general kinematics, 
in the context  
of Landau-gauge QCD with two degenerate light dynamical quarks. 
The study is based on  
the set of Schwinger-Dyson equations that govern the 
vertex form factors, derived  
within the formalism of the 
three-particle-irreducible (3PI) effective action. 
The analysis is performed 
by employing lattice data for the 
main ingredients, such as 
gluon and quark propagators,  
and three-gluon vertex.
The numerical treatment is simplified 
by decoupling the system of integral equations:
the classical form factor is determined
from a single non-linear 
equation involving only itself, while the 
remaining ones are subsequently 
computed through simple integrations.
The form factors are obtained for 
arbitrary values of space-like momenta,
and their 
angular dependence is  
examined in detail. A clear hierarchy 
is established 
at the level of the 
corresponding dimensionless effective couplings, in agreement with results of earlier 
studies. Furthermore, the 
classical form factor is found to be  
in excellent agreement with recent unquenched lattice data in the soft-gluon configuration, while 
the two non-classical dressings 
depart substantially from the lattice 
results. 
Finally, the accurate implementation of multiplicative 
renormalizability is confirmed, and the transition 
from Minkoswski to Euclidean space is elucidated. 

\end{abstract}


\maketitle

\section{Introduction}

The quark-gluon vertex, $\Gamma_{\mu}(q, p_2, -p_1)$,  
is one of the key ingredients of  
Quantum Chromodynamics (QCD)~\cite{Marciano:1977su},
playing a central role   
in the dynamical 
breaking of chiral symmetry~\cite{Nambu:1961tp,Nambu:1961fr,Lane:1974he,Politzer:1976tv} and the attendant emergence of constituent 
quark masses~\mbox{\cite{Roberts:1994dr,Maris:2003vk,Fischer:2003rp,Aguilar:2010cn,Mitter:2014wpa,Aguilar:2018epe,Gao:2021wun,Lessa:2022wqc}},  
the formation of the bound states that comprise the physical spectrum~\cite{Bender:1996bb,Maris:1999nt,Bender:2002as,Holl:2004qn,Chang:2009zb,Williams:2014iea,Williams:2015cvx,Sanchis-Alepuz:2015qra,Eichmann:2016yit,Gomez-Rocha:2015qga}, and the ongoing exploration of the phase diagram of the theory~
\cite{Roberts:2000aa,Braun:2009gm,Fukushima:2010bq,Fischer:2018sdj,Fu:2019hdw,Gao:2020qsj}. 
Given its paramount importance for
contemporary hadron physics, the quark-gluon vertex 
has been studied extensively within perturbation theory~\cite{Ball:1980ay,Kizilersu:1995iz,Davydychev:2000rt,Gracey:2014mpa,Gracey:2011vw,Bermudez:2017bpx}, by means of continuous nonperturbative approaches~\cite{Bhagwat:2004kj,LlanesEstrada:2004jz,Matevosyan:2006bk,Fischer:2006ub,Aguilar:2018epe,Qin:2013mta,Binosi:2016wcx,Hopfer:2012cnq,Rojas:2013tza,Williams:2014iea,Williams:2015cvx,Sanchis-Alepuz:2015qra,Pelaez:2015tba,Alkofer:2008tt,Mitter:2014wpa,Cyrol:2017ewj,Aguilar:2014lha,Gao:2021wun,Binosi:2016wcx,Aguilar:2016lbe,Oliveira:2018ukh,Albino:2018ncl,Tang:2019zbk,Huber:2018ned,Albino:2021rvj,Windisch:2012de, Oliveira:2020yac}, and through a plethora of lattice simulations~\cite{Skullerud:2002sk,Skullerud:2002ge,Skullerud:2003qu,Skullerud:2004gp,Lin:2005zd,Kizilersu:2021jen,Kizilersu:2006et,Sternbeck:2017ntv,Skullerud:2021pel,Oliveira:2016muq,Oliveira:2018fkj}.

An especially advantageous framework 
for studying the nonperturbative 
aspects of the quark-gluon vertex 
in the continuum is the 
formalism based on the effective 
action~\cite{Cornwall:1974vz,Cornwall:1973ts}, and particularly its 
three-particle irreducible (3PI) version, 
at the three-loop order~\cite{Berges:2004pu,Berges:2004yj,York:2012ib,Carrington:2010qq}, explored first in~\cite{Alkofer:2008tt}, and later in the  
broader study of~\cite{Williams:2015cvx}. 
One of the special characteristics of this formalism is that the resulting Schwinger-Dyson 
equation (SDE) for the quark-gluon vertex,
also known as ``equation of motion'', is composed of diagrams with all their 
fundamental vertices fully-dressed. 
This is to be contrasted with the standard 
SDE formulation, where one of the vertices 
is always kept at its classical 
(tree-level) form. 
There are two main implications stemming from this 
difference. First, the typical difficulty of dealing with diagrams multiplied by a renormalization constant (the one assigned to the vertex that 
has remained undressed) is bypassed, 
and the renormalization procedure becomes 
subtractive; for an alternative approach, see~\cite{Gao:2021wun}.
Second, the three-gluon vertex, 
which enters in the numerically dominant diagram, is 
fully-dressed; therefore, a firm grasp of its nonperturbative properties becomes indispensable 
for the successful implementation of this approach. 

In the present work we revisit the 
SDE of the {\it transversely-projected} 
quark-gluon vertex within the 
3PI approach, under the light of recent 
developments related to the 
infrared structure of the 
three-gluon vertex~\cite{Eichmann:2014xya,Ferreira:2023fva,Aguilar:2023qqd,Pinto-Gomez:2022brg,Pinto-Gomez:2024mrk}.
The importance of this special vertex for the 
dynamics of the quark-gluon vertex 
and the size of the constituent quark masses has been amply emphasized already in a series of articles~\cite{Blum:2016fib,Blum:2017uis,Alkofer:2023lrl}. Nonetheless, 
our understanding of the three-gluon vertex has advanced substantially 
in the last few years, thanks to the combined scrutiny carried out by continuum methods~\cite{Aguilar:2013vaa,Blum:2014gna,Eichmann:2014xya,Blum:2015lsa,Huber:2016tvc,Cyrol:2016tym,Corell:2018yil,Aguilar:2019jsj,Huber:2020keu,Papavassiliou:2022umz,Barrios:2022hzr,Ferreira:2023fva,Aguilar:2023qqd} and large-volume lattice simulations~\cite{Athenodorou:2016oyh,Duarte:2016ieu,Boucaud:2017obn,Vujinovic:2018nqc,Aguilar:2019uob,Aguilar:2021lke,Catumba:2021hng,Pinto-Gomez:2022brg,Pinto-Gomez:2024mrk}. The picture that 
has emerged may be summarized 
through the following key features: 
($\it a$) the form factor associated 
with the classical tensor 
displays a considerable suppression 
with respect to its tree-level value 
(unity), at intermediate and low 
momentum scales~\cite{Aguilar:2013vaa,Blum:2014gna,Eichmann:2014xya,Blum:2015lsa,Huber:2016tvc,Cyrol:2016tym,Corell:2018yil,Aguilar:2019jsj,Huber:2020keu,Papavassiliou:2022umz,Barrios:2022hzr,Ferreira:2023fva,Aguilar:2023qqd,Athenodorou:2016oyh,Duarte:2016ieu,Boucaud:2017obn,Vujinovic:2018nqc,Aguilar:2019uob,Aguilar:2021lke,Catumba:2021hng,Pinto-Gomez:2022brg,Pinto-Gomez:2024mrk}; 
($\it b$) the pivotal property of ``planar degeneracy''
reduces substantially the 
kinematic complexity of this vertex~\cite{Eichmann:2014xya,Pinto-Gomez:2022brg,Ferreira:2023fva,Aguilar:2023qqd,Pinto-Gomez:2024mrk},
furnishing simple and accurate 
forms for it, which are easily 
implementable in complicated 
computations; and  
($\it c$) 
the classical form factor diverges logarithmically in the deep infrared,
as a consequence of the 
nonperturbative masslessness of the 
ghost propagator~\cite{vonSmekal:1997ohs,Aguilar:2008xm, Fischer:2008uz, Aguilar:2013vaa,Aguilar:2021okw, Cucchieri:2007md,Cucchieri:2007rg,Bogolubsky:2007ud,Bogolubsky:2009dc,Oliveira:2009eh,Oliveira:2010xc,Cucchieri:2009zt,Boucaud:2018xup}. 

There are certain key aspects of our analysis that need to be emphasized 
from the outset. To begin with,  
a considerable simplification is implemented 
through the effective decoupling of the 
vertex SDE from the dynamical equations 
that govern the evolution of all other 
correlation functions, including the 
gap equation of the quark propagator.
In particular, 
we do not solve a system of coupled SDEs,
but use instead lattice inputs for all elements 
entering into this SDE, with the exception of the quark-gluon vertex itself. Note in particular, 
that we use lattice ingredients 
for the $N_f=2$ gluon~\cite{Ayala:2012pb,Binosi:2016xxu} and quark propagators~\cite{Oliveira:2018lln,Kizilersu:2021jen}, and 
minor variations around the best fit to the 
$N_f=2+1$ data for the 
three-gluon form factor~\cite{Aguilar:2019uob}. In addition, the system of eight 
coupled integral equations 
is simplified by retaining in their 
kernels solely the dependence on 
the classical form factor, 
$\lambda_1$, setting all others to zero.
This gives rise to a single 
self-coupled integral equation for 
$\lambda_1$; when solved, the 
$\lambda_1$ found is substituted into 
the kernels of the remaining form factors,
which are then obtained through simple
integration.

The main results  of this analysis include: 

($\it i$) The eight form factors of 
the transversely-projected 
quark-gluon vertex are computed for arbitrary values 
of the incoming space-like momenta.

($\it ii$) 
The $\lambda_1(q,p_2,-p_1)$
displays a considerable dependence on the angle between the momenta $p_1$ and $p_2$, 
while the non-classical form factors present a comparably milder dependence on this angle~\cite{Cyrol:2017ewj,Blum:2017uis}.

($\it iii$) The construction of 
a renormalization-group invariant (RGI) 
and dimensionless 
effective coupling for each form factor  
reveals a clear hierarchy amongst them, in qualitative agreement with the results of~\cite{Mitter:2014wpa,Cyrol:2017ewj,Gao:2021wun}.

($\it vi$) The infrared behaviour of $\lambda_1$ is 
particularly sensitive 
to variations of the 
gluon dressing function, 
while variations 
of the three-gluon vertex and the quark propagator have a lesser impact. 

($\it v$) The $\lambda_1$
in the soft-gluon limit 
shows excellent agreement with the lattice data of~\cite{Kizilersu:2021jen}, with a $7\%$ departure in the deep infrared, which can be further reduced through minimal  
adjustments of the three-gluon vertex dressing. However, the other two relevant form factors are very different from the lattice results.

The article is organized as follows. In Sec.~\ref{background} we 
introduce the notation and main 
ingredients, and discuss general features of 
the quark-gluon vertex. In Sec.~\ref{sdesec} 
the relevant SDE is derived, 
and its renormalization is discussed. In Sec.~\ref{numerical} we discuss the 
simplifications implemented and the inputs 
employed.
Sec.~\ref{results} contains the main results of this work; most notably, a detailed 
study of the angular dependence of the 
form factors, and a 
comparison 
with the lattice 
data of~\cite{Kizilersu:2021jen}. 
Then, in Sec.~\ref{multren} 
we verify the multiplicative renormalizability of the SDE solution for 
the classical form factor. 
In Sec.~\ref{conclusions} we present our
discussion and conclusions. Finally, in Appendix~\ref{MtoE} we elaborate on  
the transformation rules between Minkowski and Euclidean space.

\section{Notation and main ingredients}
\label{background} 

In this section, we introduce the notation and the main elements that are relevant for the ensuing considerations.
Our discussion will be restricted to the case of {\it two degenerate light dynamical quarks}, denoted by $N_f = 2$. 
Note that the formal analysis is carried out in Minkowski space;
for the purpose of the numerical treatment, some key formulas are then converted to Euclidean space, and are evaluated for space-like momenta.  

\begin{figure}[t]
    \centering
    \includegraphics[scale=0.7]{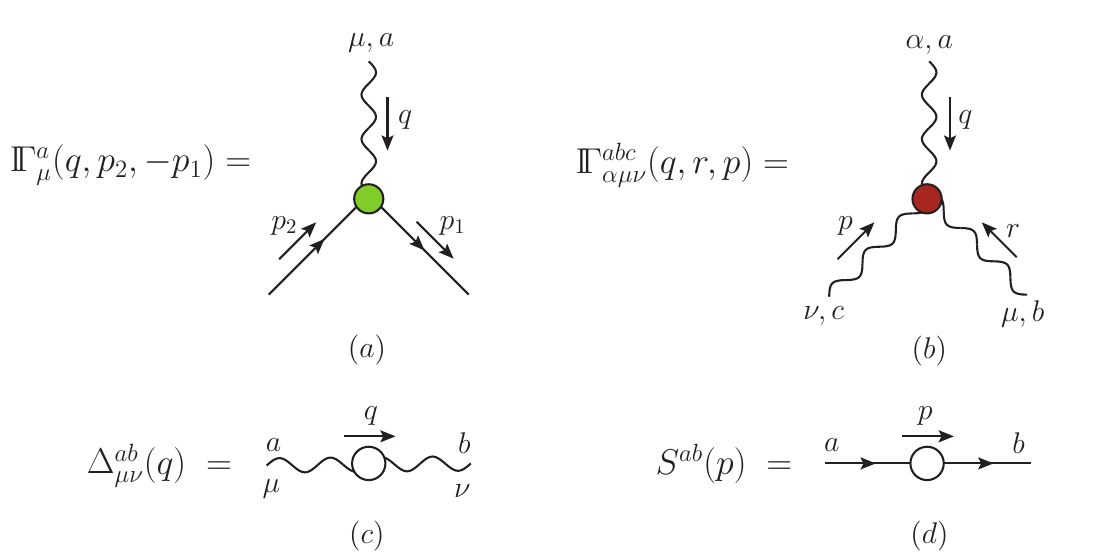}
    \caption{Diagrammatic representations of: $(a)$ the full quark-gluon vertex, $\fatg_\mu^a(q,p_2,-p_1)$, defined in Eq.~\eqref{factorG}; $(b)$ the full three-gluon vertex, $\fatg^{abc}_{\alpha\mu\nu}(q,r,p)$; $(c)$ the fully-dressed gluon propagator,  $\Delta^{ab}_{\mu\nu}(q)$; $(d)$ the full quark propagator, $S^{ab}(p)$.}
    \label{fig:prop_vert}
\end{figure}

The  full quark-gluon vertex, represented diagrammatically in panel $(a)$ of Fig.~\ref{fig:prop_vert}, is denoted by 
\begin{equation}
    \fatg_\mu^a(q,p_2,-p_1)=igt^a\fatg_\mu(q,p_2,-p_1)\,, 
    \label{factorG}
\end{equation}
where $g$ is the gauge coupling, $q$ and $p_2$ are the incoming gluon and quark momenta, \mbox{$p_1 = q + p_2$} is  the outgoing anti-quark momentum. 
In addition, \mbox{$t^a \,(a = 1, 2, \ldots, N^2-1)$} are the generators of the group SU($N$) in the fundamental representation. 
The matrices $t^a$ are hermitian and traceless, generating the closed algebra
\be
[t^a,t^b] = if^{abc}t^c\,,
\label{generators}
\ee
where $f^{abc}$ are the totally antisymmetric structure constants. 
In the case of SU$(3)$, we have that $t^a= \lambda^a/2$, where $\lambda^a$ are the Gell-Mann matrices. 
At tree-level, the  quark-gluon vertex reduces to
\begin{align}
    \Gamma_\tl^\mu(q,p_2,-p_1) = \gamma^\mu \,.
\label{qgtree}
\end{align}

The main focus of this study is the transversely-projected vertex, $\overline{\fatg}_\mu(q,p_2,-p_1)$,  defined as 
\begin{align}
     \overline{\fatg}_\mu(q,p_2,-p_1):=P_{\mu\nu}(q)\fatg^{\nu}(q,p_2,-p_1)\,, \qquad P_{\mu\nu}(q):=g_{\mu\nu}-q_\mu q_\nu/q^2\,.
\label{transvGamma}     
\end{align}
Its tree-level expression, to be denoted by $\overline{\Gamma}^\tl_\mu(q,p_2,-p_1)$, is obtained from Eq.~\eqref{transvGamma} through the substitution \mbox{$\fatg^{\nu}(q,p_2,-p_1)\to  \Gamma^{\nu}_\tl(q,p_2,-p_1)$}. 
%

In general kinematics, $\overline{\fatg}_\mu(q,p_2,-p_1)$ can be spanned by eight independent tensors, namely (Minkowski space) 
\begin{align} 
\label{decomp}
    \overline{\fatg}_\mu(q,p_2,-p_1)=\sum_{i=1}^{8}\lambda_i(q,p_2,-p_1)P_{\mu\nu}(q)\tau_{i}^\nu(p_2,-p_1) \,, 
\end{align}
where the $\lambda_i(q,p_2,-p_1)$ denote scalar form factors, which depend on three Lorentz scalars. Even though several forms for the tensors $\tau_{i}^\nu$ have been employed over the years, in the present analysis we opt for the basis put forth in~\cite{Mitter:2014wpa,Cyrol:2017ewj,Gao:2021wun};
as was shown therein, this basis has the advantages of being free of kinematic singularities and originating from gauge-invariant operators. The elements of this basis in  Minkowski space are given by 
\begin{align}   
\label{Taus}
    &\tau^\nu_{1}(p_2,-p_1) =\gamma^\nu\,, \quad &&\tau^\nu_{2}(p_2,-p_1) =  (p_1+p_2)^\nu\,, \nonumber \\
    &\tau^\nu_{3}(p_2,-p_1) = (\slashed{p}_1+\slashed{p}_2)\gamma^\nu\,, \quad &&\tau^\nu_{4}(p_2,-p_1) = (\slashed{p}_1-\slashed{p}_2)\gamma^\nu\,,\nonumber\\
    &\tau^\nu_{5}(p_2,-p_1) =  (\slashed{p}_1-\slashed{p}_2)(p_1+p_2)^\nu\,, \quad &&\tau^\nu_{6}(p_2,-p_1) =(\slashed{p}_1+\slashed{p}_2)(p_1+p_2)^\nu\,,\nonumber\\
    &\tau^\nu_{7}(p_2,-p_1) = -\frac{1}{2}[\slashed{p}_1,\slashed{p}_2]\gamma^\nu\,, \quad &&\tau^\nu_{8}(p_2,-p_1) = -\frac{1}{2}[\slashed{p}_1,\slashed{p}_2](p_1+p_2)^\nu \,.
\end{align}
It is important to emphasize that, when the $\tau_i^\nu$ of \1eq{Taus} are rotated to Euclidean space following the procedure outlined in App.~\ref{MtoE}, one recovers precisely the basis of~\cite{Mitter:2014wpa}, given in \1eq{mittbas}. This coincidence, in turn, ensures the unambiguous correspondence (and with the correct signs) between our form factors and those of~\cite{Mitter:2014wpa}.

Notice that  the full vertex, $\overline{\fatg}_\mu(q,p_2,-p_1)$, must obey the same transformation properties as the bare vertex under the charge conjugation operation $C$, namely~\cite{Kizilersu:1995iz,Davydychev:2000rt}
\begin{align}
    C\overline{\fatg}_\mu(q,p_2,-p_1)C^{-1} = -\overline{\fatg}_\mu^{T}(q,-p_1,p_2) \,.
\label{Cdressed}     
\end{align} 
Then, interchanging the momenta  \mbox{$p_1 \leftrightarrow - p_2$} in the basis defined in Eq.~\eqref{Taus}, and using the fact that 
\begin{align}
    C\gamma_{\mu}C^{-1} = -\gamma_{\mu}^{T}\,, \qquad C[\gamma_{\mu},\gamma_{\nu}]C^{-1} = [\gamma^{T}_{\mu},\gamma^{T}_{\nu}]\,,
\label{Cbare}    
\end{align}
we find that
\begin{align}
    C\tau^\nu_i(p_2,-p_1)C^{-1}&=-[\tau^{\nu}_i(-p_1,p_2)]^T\,, \qquad i=1,2,4,6,8\,, \nonumber \\
    C\tau^\nu_3(p_2,-p_1)C^{-1}&=[\tau^\nu_3(-p_1,p_2)]^T-2[\tau^\nu_2(-p_1,p_2)]^T \,,\nonumber \\
    C\tau^\nu_5(p_2,-p_1)C^{-1}&=[\tau^{\nu}_5(-p_1,p_2)]^T\,, \nonumber \\
    C\tau^\nu_7(p_2,-p_1)C^{-1}&=-[\tau^\nu_7(-p_1,p_2)]^T-[\tau^\nu_5(-p_1,p_2)]^T\,,
\label{chargecong}
\end{align}
up to terms proportional to $q^\nu$ (first equation, for $i=4$, and last one), which vanish when contracted by $P_{\mu\nu}(q)$. Therefore, in order to satisfy Eq.~\eqref{Cdressed}, we must have     
\begin{align}
\label{conjC}
   &\lambda_i(q,p_2,-p_1) = \lambda_i(q,-p_1,p_2)\,,  \quad i=1,4,6,7,8\,,\nonumber \\
   &\lambda_2(q,p_2,-p_1)+2\lambda_3(q,p_2,-p_1)=\lambda_2(q,-p_1,p_2)\,, \nonumber \\
   &\lambda_3(q,p_2,-p_1) =  - \lambda_3(q,-p_1,p_2)\,, \nonumber \\
   &\lambda_5(q,p_2,-p_1)-\lambda_7(q,p_2,-p_1)=-\lambda_5(q,-p_1,p_2)\,.
\end{align}

Combining the above relations with Lorentz invariance, which implies that the $\lambda_i$ can only depend on 
the squares of the momenta, it follows that $\lambda_3(q,p_2,-p_1) = 0$, and \mbox{$\lambda_7(q,p_2,-p_1) = 2\lambda_5(q,p_2,-p_1)$} when $p_1^2 = p_2^2$.

It is useful to separate the basis tensors into subsets that are either chirally symmetric ($cs$) or chiral symmetry breaking ($csb$): 
tensors with an odd (even) number of $\gamma$ matrices belong to the set $\tau_{cs}$ ($\tau_{csb}$). Specifically, we have 
\be
\tau_{cs} = \{\tau^\nu_{1},\tau^\nu_{5},\tau^\nu_{6},\tau^\nu_{7}\} \,, \qquad\,  
\tau_{csb} = \{\tau^\nu_{2},\tau^\nu_{3},\tau^\nu_{4},\tau^\nu_{8}\} \,.
\label{set_chiral}
\ee

In the {\it Landau gauge} that we employ, the gluon propagator,  \mbox{$\Delta^{ab}_{\mu\nu}(q)=-i\delta^{ab}\Delta_{\mu\nu}(q)$}, is fully transverse, \ie 
\begin{align}
    \Delta_{\mu\nu}(q) = \Delta(q^2)P_{\mu\nu}(q)\,, \qquad  \Delta(q^2) = \mathcal{Z}(q^2)/q^2\,,
    \label{gluoprop}
\end{align}
where $\Delta(q^2)$ denotes the scalar component of the gluon
propagator and $\mathcal{Z}(q^2)$ the 
corresponding dressing function. 
The diagrammatic representation of $\Delta_{\mu\nu}(q)$ 
is given in panel $(c)$ of Fig.~\ref{fig:prop_vert}.

In addition, we denote by \mbox{$S^{ab}(p)=i\delta^{ab}S(p)$} the quark propagator [see panel $(d)$ of Fig.~\ref{fig:prop_vert}], whose standard decomposition is given by
\begin{align} 
\label{qprop}
    S^{-1}(p)=A(p^2)\slashed{p}-B(p^2)\,,
\end{align}
where $A(p^2)$ and $B(p^2)$ are the Dirac vector and scalar components, respectively. 
The dynamical quark mass function, 
${\mathcal M}(p^2)$,  
is then defined as \mbox{${\mathcal M}(p^2) = B(p^2)/ A(p^2)$}.
At tree-level, 
\be
S_0^{-1}(p)= \slashed{p} - m_q \,, 
\label{S0}
\ee
such that $A_0 = 1$ and $B_0 = m_q$, where $m_q$ is the current quark mass.

Finally, we introduce 
the three-gluon vertex, $\fatg_{\alpha\mu\nu}^{abc}(q,r,p)=gf^{abc}\fatg_{\alpha\mu\nu}(q,r,p)$, depicted in panel $(b)$ of Fig.~\ref{fig:prop_vert}.
At tree-level, $\fatg_{\alpha\mu\nu}(q,r,p)$ reduces 
to the standard expression
\begin{align} 
\label{3g0}
    \Gamma_\tl^{\alpha\mu\nu}(q,r,p)=g^{\mu\nu}(r-p)^\alpha +g^{\alpha\nu}(p-q)^\mu+g^{\alpha\mu}(q-r)^\nu\,.
\end{align}
Note that, in our analysis, 
the three-gluon 
vertex is naturally contracted by three transverse projectors, namely 
\begin{align}
    \overline{\fatg}_{\alpha\mu\nu}(q,r,p)=P_\alpha^{\alpha^\prime}(q)P_\mu^{\mu^\prime}(r)P_\nu^{\nu^\prime}(p)\fatg_{\alpha^\prime\mu^\prime\nu^\prime}(q,r,p)\,.
\end{align}

\section{SDE of the 
quark-gluon vertex} \label{sdesec}

In this section we introduce the SDE of the quark-gluon vertex that will be employed in our analysis, and elaborate on the procedure adopted 
for its renormalization. 

\subsection{General structure} \label{sdegen}

In this study, we employ the formulation of the quark-gluon  SDE  derived within the framework of the 3PI effective action~\cite{Cornwall:1974vz,Cornwall:1973ts}, at the {\it three-loop} level~\cite{Berges:2004pu,Berges:2004yj,Carrington:2010qq,York:2012ib,Alkofer:2008tt,Williams:2015cvx}. 
%
\begin{figure}[t]
    \centering
    \includegraphics[scale=0.6]{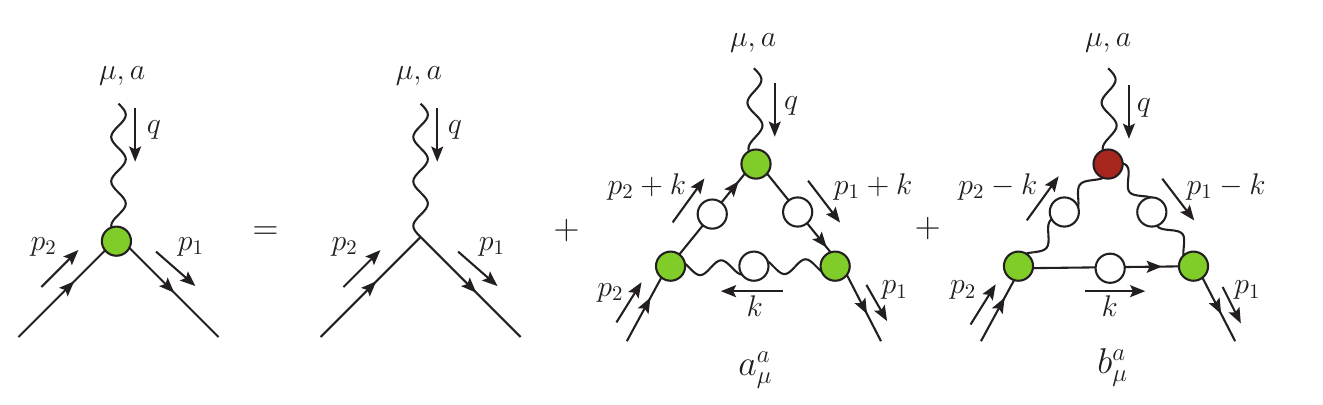}
    \caption{ Diagrammatic representation of the SDE for the full quark-gluon vertex, ${\fatg}_\mu^a(q,p_2,-p_1)$,  derived from the 3PI effective action at the three-loop level. White circles denote full propagators,  and the green (red) circles denote the fully dressed quark-gluon (three-gluon)-vertices.  Diagrams $a_\mu$ and $b_\mu$  are often referred to as ``Abelian'' and ``non-Abelian'', respectively.}
    \label{fig:sde}
\end{figure}

It is well-known that, within the $n$PI formalism, 
the SDE of a given Green's function (also known as ``equations of motion'') is obtained through the 
functional differentiation and subsequent extremization of the effective action with respect to the Green's function in question. 
In the 3PI case, the relevant  
SDEs are derived from the effective action shown 
diagrammatically in Figs.~1-2 of \cite{Berges:2004pu,Berges:2004yj,Williams:2015cvx}. At this level of approximation, all propagators (gluon, ghost, and quark)
comprising this action are fully dressed, and so are 
the quark-gluon and three-gluon vertices; instead,  
the four-gluon vertex is kept at its tree-level value. 
Note also that the resulting propagator SDEs
are ``two-loop dressed'', while the vertex SDEs are 
``one-loop dressed''. 

Specializing to the quark-gluon vertex,
the corresponding SDE is diagrammatically depicted 
in Fig.~\ref{fig:sde}. In particular, 
the transversely-projected quark-gluon vertex can be expressed in terms of the Abelian  $(a^\mu)$  and non-Abelian $(b^\mu)$ diagrams as 
\begin{align}   
\label{quark_sde}
    \overline{\fatg}^\mu(q,p_2,-p_1)=  \overline{\Gamma}_\tl^{\mu}(q,p_2,-p_1) +  \overline{a}^\mu (q,p_2,-p_1) +  \overline{b}^{\mu} (q,p_2,-p_1) \,,
\end{align}
where 
\begin{align}
    \overline{a}^\mu (q,p_2,-p_1) = P_\nu^\mu(q) a^\nu (q,p_2,-p_1) \,, \quad
    \overline{b}^\mu (q,p_2,-p_1) = P_\nu^\mu(q) b^\nu (q,p_2,-p_1) \,.
\end{align}

After carrying out the color algebra, the corresponding  contributions in Minkowski space
are given by
\begin{align}   
\label{diagrams}
    \overline{a}_\mu (q,p_2,-p_1)\!= &\kappa_a \!\! \int_k \! \!\Delta(k^2) \overline{\fatg}^\alpha(-k,k_1,-p_1)S(k_1) \overline{\fatg}_\mu(q,k_2,-k_1) S(k_2) \overline{\fatg}_\alpha(k,p_2,-k_2)\,, \nonumber \\
    \overline{b}_\mu (q,p_2,-p_1)\!= &\kappa_b \!\!\int_k\!\! \Delta(\ell_1^2)\Delta(\ell_2^2)\overline{\fatg}_{\mu\alpha\beta}(q,\ell_1,-\ell_2)  \overline{\fatg}^\alpha(-\ell_1,k,-p_1) S(k) \overline{\fatg}^\beta(\ell_2,p_2,-k) \,,
\end{align}
where \mbox{$k_1 := k+p_1$}, \mbox{$k_2 := k+p_2$}, \mbox{$\ell_1 := k-p_1$}, and \mbox{$\ell_2 := k-p_2$}, and we have introduced the  factors
\begin{align}
    \kappa_a:=-ig^2\left(C_{\rm F}-\frac{C_{\rm A}}{2}\right) \,, \qquad \kappa_b := \frac{ig^2C_{\rm A}}{2}\,,
\end{align}
where  $C_{\rm F}$ and $C_\mathrm{A}$ denote the eigenvalues of the Casimir operator in the fundamental and adjoint representations, respectively [$C_{\rm F} = (N^2-1)/2N$ and  $C_\mathrm{A} = N$ for SU($N$)]. 

Finally, we denote by 
\be\label{eq:int_measure}
\int_{k} := \frac{1}{(2\pi)^4} \int \!\!{\rm d}^4 k \,,
\ee
the integration over virtual momenta, 
where the use of a symmetry-preserving regularization scheme is implicitly assumed.

It is straightforward to show that the expressions in \1eq{diagrams} preserve charge conjugation symmetry; therefore, the form factors derived from these integrals automatically satisfy the constraints given in \1eq{conjC}.

In order to derive the dynamical equations governing the form factors $\lambda_i(q,p_2,-p_1)$ in general kinematics, one has to construct a set of appropriate projectors, ${\mathcal P}_{i}^{\mu}(q,p_2,-p_1)$, satisfying the basic property
\begin{equation}
    \text{Tr} \left[{\mathcal P}^\mu_{i}(q,p_2,-p_1) P_{\mu\nu}(q)\tau^\nu_j(p_2,-p_1)\right] = \delta_{ij}\,,
\label{projprop}
\end{equation}
such that 
\begin{align}
    \label{formfactors}
    \lambda_i(q,p_2,-p_1) =   \text{Tr} \left[{\mathcal P}_{i\mu}(q,p_2,-p_1)\overline{\fatg}^\mu_{\s R} (q,p_2,-p_1) \right]\,.
\end{align}
The explicit construction of the 
projectors ${\mathcal P}_{i}^\mu$
proceeds by casting them in the form 
\begin{equation}
    {\mathcal P}_{i}^\mu(q,p_2,-p_1) = \sum_{j=1}^8 C_{ij}(p_2,-p_1)
    \tau_{j}^\mu(p_2,-p_1)\,,
\end{equation}
imposing \1eq{projprop}, and solving the 
resulting system to determine the coefficients $C_{ij}$. This procedure yields 
\begin{align}  
\label{projectors}
    {\mathcal P}_{1}^{\mu} &\!=\!c_1\!\left[4h 
    \tau^\mu_{1}+r^2 \tau^\mu_{5}-q^2\tau^\mu_{6}\right]\,, 
    \, && {\mathcal P}_{5}^{\mu} \!=\!c_2\!\left[4h(r^2\tau^\mu_{1}\!-\!2\tau^\mu_{7})\!+\!(3r^4\!+\!4h)\tau^\mu_{5}\!-\!3r^2q^2\tau^\mu_{6}\right]\,,\nonumber\\
     {\mathcal P}_{2}^{\mu} &\!=\! c_1\!\left[q^2(\tau^\mu_{2}\!+\!\tau^\mu_{3})-r^2\tau^\mu_{4}\right]\,, 
    \, && {\mathcal P}_{6}^{\mu} \!=\!-c_2\!\left[4hq^2 \tau^\mu_{1}+3q^2(r^2\tau^\mu_{5}\!-\!q^2\tau^\mu_{6})\right]\,,\nonumber\\
    \ {\mathcal P}_{3}^{\mu} &\!=\! c_1\!\left[q^2 (\tau^\mu_{2}\!-\!\tau^\mu_{3})+r^2\tau^\mu_{4}\right] \,,
    \, && {\mathcal P}_{7}^{\mu} \!=\!-2c_1\!\left[\tau^\mu_{5}+2\tau^\mu_{7}\right]\,,\nonumber\\
     {\mathcal P}_{4}^{\mu} &\!=\! c_1\!\left[r^2 (\tau^\mu_{3}\!-\!\tau^\mu_{2})\!-\!(p_1\!+\!p_2)^2\tau^\mu_{4}\!-\!2\tau^\mu_{8}\right]\,,
    && {\mathcal P}_{8}^{\mu} \!=\!-4c_2\!\left[2h \tau^\mu_{4}+3q^2\tau^\mu_{8}\right]\,, 
\end{align}
where we suppressed the argument $(q,p_2,-p_1)$ in the ${\mathcal P}_{i}^{\mu}$ and $\tau_{i}^\mu$, and introduced the definitions 
\be
r^2:=p_1^2-p_2^2 \,,
\qquad
h := p_1^2p_2^2 - (p_1\cdot p_2)^2\,,
\qquad
c_1:=1/32h\,,
\qquad
c_2:=1/128h^2 \,.
\label{somedef}
\ee

The SDE of \1eq{quark_sde} may be expressed in terms of the  
individual $\lambda_i(q,p_2,-p_1)$. To that end, 
it is convenient to denote by ${\mathbb A}_{i}$ and ${\mathbb B}_{i}$ the contributions arising from the contraction of 
the Abelian and non-Abelian diagrams in  Fig.~\eqref{fig:sde} 
by the projectors ${\mathcal P}_{i}^{\mu}$, namely
\begin{align}
    \label{formfactors2}
    {\mathbb A}_{i}(q,p_2,-p_1) & :=   \text{Tr} \Big[{\mathcal P}_{i\mu}(q,p_2,-p_1)\, \overline{a}^{\mu} (q,p_2,-p_1)  \Big]\,,\nonumber \\ 
    {\mathbb B}_{i}(q,p_2,-p_1) & :=   \text{Tr} \Big[{\mathcal P}_{i\mu}(q,p_2,-p_1)\, \overline{b}^{\mu} (q,p_2,-p_1)  \Big]\,.
\end{align}
Then, using \2eqs{projprop}{formfactors} it is easy to arrive at the system 
\begin{align}   
    &\lambda_i(q,p_2,-p_1)=\delta_{i1}+ {\mathbb A}_{i}(q,p_2,-p_1)+ {\mathbb B}_{i}(q,p_2,-p_1) \,, \qquad i=1,\ldots,8 \,.
    \label{lambdai} 
\end{align}

\subsection{Renormalization}
\label{sderen}
We next turn to the renormalization of the SDE for the quark-gluon vertex. Clearly, all quantities appearing 
in \1eq{quark_sde} are bare; the conversion to their renormalized counterparts is carried out multiplicatively,  using the standard relations~\cite{Marciano:1977su}
\begin{align} 
\label{renconst}
    &\Delta_{\s R}(q^2)= Z^{-1}_{A} \Delta(q^2)\,,  \qquad S_{\s R}(p)= Z^{-1}_{F} S(p)\,,\qquad g_{\s R} = Z_g^{-1} g\,, \nonumber\\
    &\fatg^{\alpha\mu\nu}_{\!\!\s R}(q,r,p) =  Z_3 \fatg^{\alpha\mu\nu}(q,r,p)\,,  \qquad \fatg^{\mu}_{\!\!\s R}(q,p_2,-p_1) = Z_1 \fatg^{\mu}(q,p_2,-p_1)\,,
\end{align} 
where the subscript ``$R$'' denotes renormalized quantities, and  $Z_{A}$, $Z_{F}$, $Z_g$, $Z_{3}$, and $Z_{1}$ are the corresponding (cutoff-dependent) renormalization constants.
In addition, we employ the exact relations 
\be
    Z_g^{-1} = Z_1^{-1} Z_A^{1/2} Z_F\, = Z_3^{-1} Z_A^{3/2} \,,
\label{eq:sti_renorm}
\ee 
which are imposed by the fundamental Slavnov-Taylor identities (STIs)~\cite{Taylor:1971ff,Slavnov:1972fg}. 

We point out that the renormalization constant of the quark current mass, usually denoted by $Z_{m_q}$, has been omitted in 
\1eq{renconst}. 
To be sure, in a self-contained SDE analysis, where the quark propagator is  determined by its own SDE, 
the inclusion of $Z_{m_q}$ would be  indispensable.
Indeed, in such a treatment, the current mass would appear explicitly in the quark SDE through the tree-level quark propagator, see~\1eq{S0}; 
then, upon renormalization, $Z_{m_q}$ would 
enter in the system of SDEs.
Instead, in the present work we will employ renormalized lattice inputs for the external ingredients of \1eq{quark_sde}. Then, since \1eq{quark_sde} does not explicitly contain the quark current mass, its corresponding renormalization constant does not appear in our analysis. For further details on this subject, 
the reader is referred to~\cite{Fischer:2006ub,Mitter:2014wpa,Eichmann:2016yit,Cyrol:2017ewj,  Fischer:2018sdj,Fu:2019hdw, Aoki:2019cca,Gao:2020qsj,Gao:2020fbl,Dupuis:2020fhh}, and references therein.

By substituting the relations of \1eq{renconst} into \eqref{quark_sde}, and using \1eq{eq:sti_renorm}, we readily obtain the renormalized version of \eqref{quark_sde}, expressed as 
\begin{align}   
\label{quark_sde_r}
    \overline{\fatg}^{\mu}_{\!\s R}(q,p_2,-p_1)=   Z_1\overline{\Gamma}_\tl^{\mu}(q,p_2,-p_1) +  \overline{a}^{\mu}_{\s R} (q,p_2,-p_1) +  \overline{b}^{\mu}_{\s R} (q,p_2,-p_1) \,,
\end{align}
where the subscript ``$R$'' in $\overline{a}^{\mu}_{\s R}$ and $\overline{b}^{\mu}_{\s R}$ indicates that the expressions provided in \1eq{diagrams} have been replaced by their renormalized counterparts, as defined in \1eq{renconst}. 
Then, from \1eq{quark_sde_r} one may readily derive the renormalized analog of \1eq{lambdai}, namely  
\begin{align}   
    &\lambda_{i,\s R}(q,p_2,-p_1)= Z_1 \delta_{i1}+ {\mathbb A}_{i,\s R}(q,p_2,-p_1)+ {\mathbb B}_{i,\s R}(q,p_2,-p_1) \,, \qquad i=1,\ldots,8 \,.
    \label{lambdaiRZ} 
\end{align}

Note that, 
due to the fact that 
all vertices comprising the diagrams $\overline{a}^\mu$ and $\overline{b}^\mu$ 
are fully-dressed, no renormalization 
constants appear multiplying them in \1eq{quark_sde_r}. In fact, 
the only renormalization constant that survives 
in \1eq{quark_sde_r}, 
namely $Z_1$, is multiplying 
the tree-level 
contribution, thus converting the procedure 
of renormalization into subtractive instead of 
multiplicative. 
This is one of the main advantages 
offered by the 3PI formulation~\mbox{\cite{Carrington:2010qq,York:2012ib,Alkofer:2008tt,Williams:2015cvx}}, bringing about a major operational simplification.

To determine $Z_1$  we employ a variation of the momentum subtraction (MOM) scheme \mbox{\cite{Celmaster:1979km,Hasenfratz:1980kn}}, the so-called \MOMt{} scheme~\cite{Skullerud:2002ge}.
In particular, 
denoting 
by $\lambda_1^\sg(p^2):=\lambda_1(0,p,-p)$
the classical form factor of the quark-gluon vertex in the soft-gluon limit ($q \to 0$),
this particular scheme is defined by the prescriptions
\be 
\Delta^{-1}_{\s R}(\mu^2) = \mu^2 \,, \qquad A_{\s R}(\mu^2) = 1  \,, \qquad \lambda_{1, \s R}^\sg(\mu^2)=1\,.
\label{ren_conds}
\ee

The implementation of this condition at the 
level of \1eq{lambdaiRZ} proceeds 
by considering the case $i=1$ 
and taking the limit $q \to 0$. 
Employing the notation 
\mbox{${\mathbb A}_{i,\s R}(0,p,-p) := {\mathbb A}^\sg_{i,\s R}(p^2)$} and \mbox{${\mathbb B}_{i,\s R}(0,p,-p) := {\mathbb B}^\sg_{i,\s R}(p^2)$}, we find 
\begin{align}
    \lambda_{1,\s R}^\sg(p^2) &= Z_1 + {\mathbb A}_{1,\s R}^\sg(p^2) + {\mathbb B}_{1,\s R}^{\,\sg}(p^2)\,.
    \label{lambda1_re}
\end{align}
Then, by imposing the renormalization condition of \1eq{ren_conds}, 
we find  
\begin{align}
   Z_{1} = 1-
   {\mathbb A}_{1,\s R}^\sg(\mu^2) - {\mathbb B}_{1,\s R}^{\,\sg}(\mu^2)
   \,. 
\label{Z1_exp}    
\end{align}
Thus, substituting Eq.~\eqref{Z1_exp}
into \1eq{lambdaiRZ}, we arrive at the renormalized version of \1eq{lambdaiRZ} , namely  
%
\begin{align}    \label{lambdaiR}
    &\lambda_{i,\s R}(q,p_2,-p_1)=[1-{\mathbb A}_{i,\s R}^\sg(\mu^2) - {\mathbb B}_{i,\s R}^{\,\sg}(\mu^2)
    ]\delta_{i1}+ {\mathbb A}_{i,\s R}(q,p_2,-p_1)+{\mathbb B}_{i,\s R}(q,p_2,-p_1) \,.
\end{align}
\noindent From now on the index ``$R$'' 
will be suppressed, to avoid notational clutter. 

The SDEs in \1eq{lambdaiR} will be solved under certain simplifying assumptions that we discuss in detail in the next section.

\section{Numerical setup and inputs}  \label{numerical}

In this section we discuss the basic simplifications 
imposed on \1eq{lambdaiR}, 
derive its Euclidean version,  
and provide inputs for the  
propagators and vertices that 
comprise the terms ${\mathbb A}_{i}$ 
and ${\mathbb B}_{i}$. 

\subsection{Simplifications}  \label{simp}

To reduce the algebraic complexity of this system, we approximate all transversely-projected quark-gluon vertices appearing on the rhs of Eq.~\eqref{diagrams} by retaining only their classical tensorial structures. Specifically, we set
 \begin{align} \label{aprox1}
    \overline{\fatg}_\mu(q,p_2,-p_1) \to \lambda_1(q,p_2,-p_1)P_{\mu\nu}(q)\gamma^\nu \,.
\end{align}

Notice that implementing this approximation leads to two major simplifications: {\it (i)} the dynamical equation for the classical form factor, $\lambda_1(q,p_2,-p_1)$, decouples from the seven remaining form-factors, and {\it (ii)} the equation for the remaining form factors, $\lambda_i(q,p_2,-p_1)$ for $i\neq1$, are expressed in terms of only $\lambda_1$. Therefore, $\lambda_1(q,p_2,-p_1)$ is described by an integral equation, whereas the remaining $\lambda_i$ are expressed in terms of ordinary integrals involving the form factor $\lambda_1$.

Regarding the three-gluon vertex, 
we retain only its tree-level tensorial structure, and resort to 
the planar degeneracy approximation 
for the associated form factor, which has been validated by a series of studies~\cite{Eichmann:2014xya,Pinto-Gomez:2022brg,Ferreira:2023fva,Aguilar:2023qqd,Pinto-Gomez:2024mrk}.
Specifically, 
$\overline{\fatg}^{\,\mu\alpha\beta}(q,\ell_1,-\ell_2)$ can be accurately approximated by the compact form
\be 
\overline{\fatg}^{\mu \alpha \beta }(q,\ell_1,-\ell_2) =\Ls(s^2) \overline{\g}_{\!0}^{\, \mu \alpha \beta}(q,\ell_1,-\ell_2) \,, \qquad s^2= \frac{1}{2}(q^2+\ell_1^2+\ell_2^2)\,,
\label{compact}
\ee
where $\overline{\g}_{\!0}^{\, \mu \alpha \beta}(q,\ell_1,-\ell_2) =  P_{\mu'}^\mu(q)  P_{\alpha'}^{\alpha}(\ell_1)  P_{\beta'}^\beta(\ell_2) 
\g_{\!0}^{\,\mu'\! \alpha' \!\beta'}(q,\ell_1,-\ell_2)$,  with $\g_{\!0}^{\mu'\! \alpha'\! \beta'}$ denoting the three-gluon vertex at tree-level given by Eq.~\eqref{3g0}. The function $\Ls(s^{2})$ is  the form factor associated with the soft-gluon limit of the three-gluon
vertex, $(q=0, \, \ell_1 =\ell_2)$, and has been  determined from various lattice simulations~\mbox{\cite{Athenodorou:2016oyh,Duarte:2016ieu,Boucaud:2017obn, Vujinovic:2018nqc,Pinto-Gomez:2022brg,Aguilar:2019uob,Aguilar:2021lke,Aguilar:2021okw,Pinto-Gomez:2024mrk}}.

\subsection{Euclidean space}  \label{Euc}

The conversion of \1eq{lambdaiR} to Euclidean space proceeds by assuming that the external momenta are
space-like, \eg $q^2 \to -q^2_{\s E}$ with $q^2_{\s E} \geq 0$, and similarly for $p_1$ and $p_2$.
We then apply the following standard conversion rules 
\begin{align}
    \Delta_\euc(q_\euc^2) = - \Delta(-q_\euc^2), \quad A_\euc(q_\euc^2) = A(-q_\euc^2), \quad B_\euc(q_\euc^2) = B(-q_\euc^2),  \quad 
    L^\euc_\sg(q_\euc^2) =  L_\sg(-q_\euc^2)\,, 
\end{align}
and suppress the index ``E'' for simplicity.

In addition, it is convenient 
to employ 
hyperspherical coordinates, 
writing the integral measure as 
\begin{align}
    &\int_{k} \to \int_{\rm{E}} =\frac{i}{16\pi^{3}}\!\!\int_{0}^{\infty}\!\!\!\!\!\! dz\, z   \!\!\int_{0}^{\pi} \!\!\!\!\!d \phi_{1} s^{2}_{\phi_{1}} \!\!\int_{0}^{\pi}\!\!\!\!\! d \phi_{2} s_{\phi_{2}} \,,
\end{align}
where $s_{\phi_1} :=\sin{\phi_1}$. We then express all relevant form factors as functions 
of $p_1^2$, $p_2^2$, and the angle between them, $\theta$, namely
\be 
\lambda_i(q,p_2,-p_1) \to \lambda_i(p_1^2,p_2^2,\theta)\,.
\label{conv}
\ee
After implementing the above steps, we 
find that \1eq{lambdaiRZ} 
assumes the schematic form 
\begin{align}   
\lambda_i(p_1^2,p_2^2,\theta)=Z_1\delta_{i1}&+\int_{\rm{E}} \!\! {\mathcal{K}}_{i\mathbb A}\lambda_1^3   
    + \!\int_{\rm{E}}\!\! {\mathcal{K}}_{i\mathbb B}\lambda_1^2 \,, 
    \label{lambdaieuc2}
\end{align}
where ${\mathcal{K}}_{i\mathbb A}$ 
and ${\mathcal{K}}_{i\mathbb B}$ 
 are the kernels
 of diagrams $a_{\mu}$ and $b_{\mu}$ in \fig{fig:sde}, respectively.

\begin{figure}[t]
    \centering
    \setbox1=\hbox{\includegraphics[height=5cm]{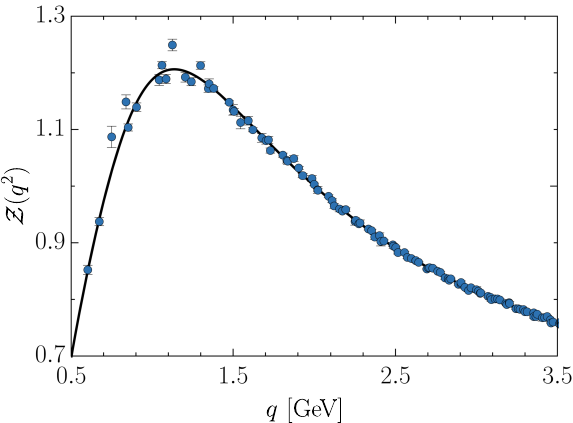} \hfil}
    \includegraphics[scale=0.7]{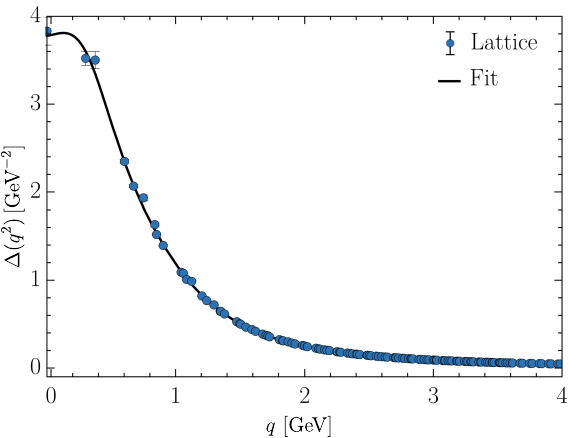}
    \hspace{-0.5cm}\llap{\makebox[\wd1][r]{\raisebox{1.5cm}{\includegraphics[scale=0.37]{fig3a}}}}\hfil
    \includegraphics[scale=0.74]{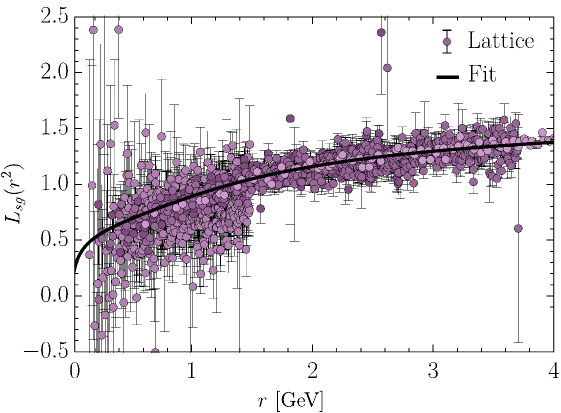} 
    \caption{\!\!{\it Left panel:} The lattice data of~\cite{Ayala:2012pb,Binosi:2016xxu} for the gluon propagator, $\Delta(q^2)$, with  $N_f = 2$ (points), together with the fit given by  Eq.~(A1) of~\cite{Aguilar:2023mam} (black solid line). In the inset we show the gluon dressing function, ${\mathcal Z}(q^2)$, defined in Eq.~\eqref{gluoprop}. {\it Right panel:} Lattice data 
    of~\cite{Aguilar:2019uob} 
    for $\Ls(r^{2})$ with $N_f=2+1$ (points), converted to the \MOMt{} scheme, which entails a rescaling by a factor of 1.16~\cite{Aguilar:2023mam}. The black continuous curve is the corresponding fit, given by Eq.~(A1) of~\cite{Aguilar:2023mam}.}
    \label{inputsDL}
\end{figure}

Moreover, 
in conformity with 
\1eq{Z1_exp}, 
the renormalization constant $Z_1$ is
obtained as the 
soft-gluon limit 
of Eq.~\eqref{lambdaieuc2} for $i=1$; 
specifically, 
setting 
$q=0$, $p_1=p_2 :=p$, and $\theta=0$, and subsequently fixing 
$p^2=\mu^2$, 
we find 
\begin{align}   
Z_1 = 1 - \lim_{q\to 0}\left[ \int_{\rm{E}} \!\! {\mathcal{K}}_{1\mathbb A}\lambda_1^3   
    + \!\int_{\rm{E}}\!\! {\mathcal{K}}_{1\mathbb B}\lambda_1^2
\right]_{p^2=\mu^2}\,.
\label{Z1final}
\end{align}

\subsection{Inputs}  \label{inp}

In what follows, the system of equations for the  $\lambda_i$ formed by Eqs.~\eqref{lambdaieuc2} and \eqref{Z1final}  are solved treating $\Delta(q^2)$, $A(p^2)$, ${\mathcal M}(p^2)$, and $\Ls(r^2)$ as external inputs. 

{\it (i)} For $\Delta(q^2)$ we use a fit to the lattice data from~\cite{Ayala:2012pb,Binosi:2016xxu}, with  $N_f = 2$ (two degenerate light quarks), computed with current masses between $20$ to $50$~MeV, and pion masses ranging from $270$ to $510$~MeV. The functional form of this fit is provided in Eq.~(A1) of~\cite{Aguilar:2023mam}, and is shown in the left panel of Fig.~\ref{inputsDL},
together with 
the corresponding 
dressing function $\mathcal{Z}(q^2)$. 
 
{\it (ii)}  
When dealing with 
the form factor 
$\Ls(r^2)$, additional care is needed, because, at present, 
there are no available lattice 
data for this quantity with 
\mbox{$N_f = 2$}. 
Given this limitation, we will 
employ a fit for lattice data 
with \mbox{$N_f = 2+1$}, 
which was used 
recently in the analysis of~\cite{Aguilar:2023mam}, see Eq.~(A1) therein. Note that this simulation
involves 
two light quarks with a current mass of $1.3$~MeV, and a heavier one with a current mass of $63$~MeV~\cite{Aguilar:2019uob}. Moreover, the lattice results of~\cite{Aguilar:2019uob} were originally computed in the so-called asymmetric MOM scheme, defined by the prescription $\Ls^{as}(\mu^2) = 1$. To employ them in our analysis, we convert them to the \MOMt{} scheme through the rescaling $\Ls(r^2) = 1.16\Ls^{as}(r^2)$, valid for $\mu = 2$~GeV, as determined in the Appendix B of~\cite{Aguilar:2023mam}.

In the right panel of Fig.~\ref{inputsDL} we show the \MOMt{} converted lattice data and the corresponding 
fit (black curve), denoted by 
$\Ls^{*}(r^2)$, which minimizes the $\chi^2$ deviation from the data. The curve 
$\Ls^{*}(r^2)$ will serve as our reference input for 
$\Ls(r^2)$, \ie 
$\Ls(r^2) \to \Ls^{*}(r^2)$, 
for the bulk of our computations. However, minor variations from  $\Ls^{*}(r^2)$ 
will be implemented in order to 
optimize the coincidence with 
the lattice data for $\lambda_1$ 
in the soft-gluon limit, 
and check the numerical stability of the entire procedure.

\begin{figure}[t]
    \includegraphics[scale=0.74]{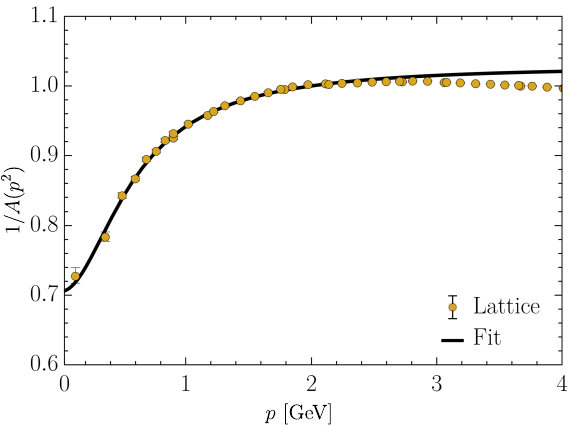} \hfil
    \includegraphics[scale=0.74]{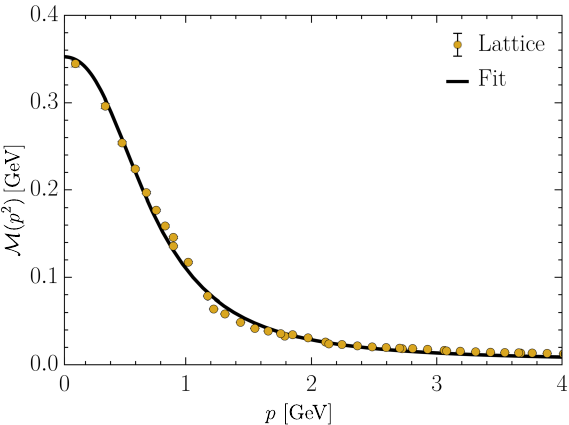} \hfil
    \caption{The lattice data (points) of~\cite{Oliveira:2018lln,Kizilersu:2021jen} for the quark wave function, $1/A(p^2)$
    (left panel) and the running mass, ${\mathcal M}(p^2)$ (right panel), both corresponding to the ``L08'' setup. The fits (black solid lines) given by Eqs.~(A4) and (A5) of~\cite{Aguilar:2023mam}
    are also shown.}
    \label{inputsAM}
\end{figure}

{\it (iii)} For the quark wave function, $1/A(p^2)$,  and the corresponding dynamical  mass,  ${\mathcal M}(p^2)$, we employ fits for the setup  denominated  ``L08''   
in the lattice simulation of~\cite{Oliveira:2018lln,Kizilersu:2021jen}. This simulation was performed for a current
quark mass \mbox{$m_q =6.2$~MeV}, and a pion mass \mbox{$m_\pi =280$~MeV}. 
The 
functional forms of the fits for $A(p^2)$ and   ${\mathcal M}(p^2)$ are given in Eqs.~(A5) and~(A6) 
of~\cite{Aguilar:2023mam}, respectively; these functional forms were 
adjusted to remove certain 
lattice artifacts in the ultraviolet, and to reproduce the respective one-loop resumed perturbative behaviour.  Both the lattice data and the corresponding fits for $A(p^2)$ and   ${\mathcal M}(p^2)$ are shown in Fig.~\ref{inputsAM}.

{\it (iv)} All these inputs are renormalized in the \MOMt{} scheme defined by Eq.~\eqref{ren_conds}, at the renormalization point \mbox{$\mu= 2$~GeV}. For this particular $\mu$ we 
need to choose a value for 
$\alpha_s(\mu^2) := g^2(\mu^2)/4\pi$; 
whereas in~\cite{Aguilar:2023mam}
the estimate \mbox{$\alpha_s(\mu^2) = 0.47$} was obtained 
by combining  one-loop calculations and fits to lattice data, our SDE 
reproduces the lattice results for values of
the strong charge 
in the vicinity of 
\mbox{$\alpha_s(\mu^2) = 0.55$} (see next section).

\section{Results}  
\label{results}

We next solve \1eq{lambdaieuc2} for $i=1$ iteratively, 
in order to obtain $\lambda_1$.
The integration is performed using a double-precision adaptive routine based on the Gauss-Kronrod integration rule~\cite{Berntsen:1991:ADA:210232.210234}. The external momenta grid interval ranges logarithmically from $[10^{-3}, 10^3]~\mbox{GeV}^2$, with 30 points, while the external angle grid is uniformly distributed across 10 points within the range $[0,\pi]$. The interpolations in three variables, needed for evaluating the $\lambda_1$, are performed with B-splines~\cite{de2001practical}.
Once $\lambda_1$ has been determined, we substitute it into \1eq{lambdaieuc2} for $i=2,\cdots,8$, and obtain all remaining $\lambda_i$ through simple integration. 
In the rest of this Section, we present the 
main results of this analysis.

\subsection{Classical (tree-level) form factor} \label{sec:classical}

\begin{figure}[ht]
    \centering
    \includegraphics[scale=0.8]{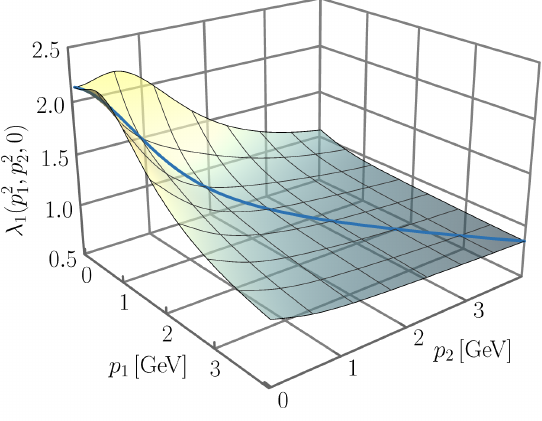}
    \hspace{0.5cm}
    \includegraphics[scale=0.8]{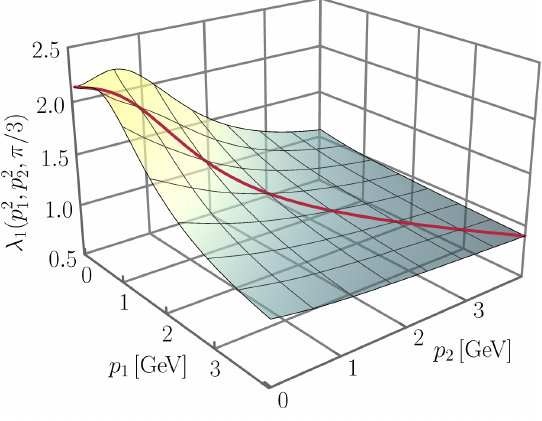}
    \includegraphics[scale=0.8]{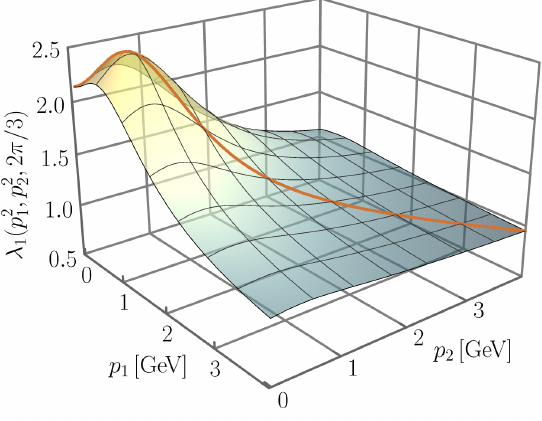} 
    \hspace{0.5cm}
    \includegraphics[scale=0.8]{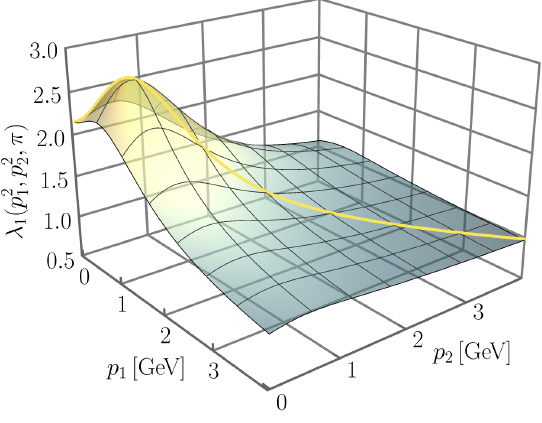}
    \caption{The form factor $\lambda_1(p_1^2, p_2^2, \theta)$ plotted as a function of the antiquark, $p_1$, and quark, $p_2$, momenta, for  fixed angles $\theta=0$ (upper left), $\theta=\pi/3$ (upper right), $\theta=2\pi/3$  (lower left),   and $\theta=\pi$ (lower right). Four kinematic limits are highlighted in the diagonals of each of the 3D surfaces:  soft-gluon (blue), totally symmetric (red), quark-symmetric (brown), and asymmetric (yellow) configurations.}
    \label{figlambda1kin}
\end{figure}

\begin{figure}[ht]
    \centering
    \includegraphics[scale=0.8]{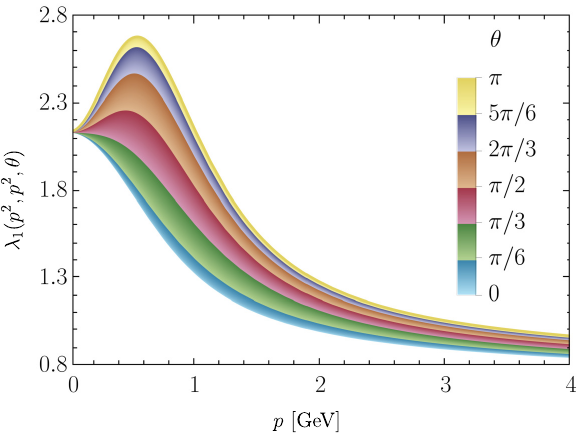}
    \hspace{0.5cm}
    \includegraphics[scale=0.8]{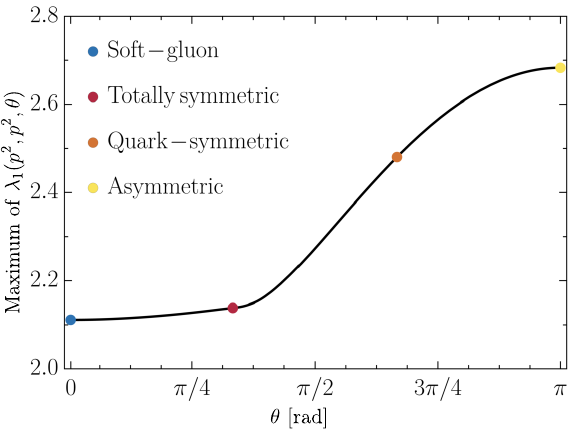}
    \caption{\emph{Left panel:} The  dependence of $\lambda_1(p_1^2,p_2^2,\theta)$ on the angle $\theta$, when we set $p^2_1=p^2_2=p^2$. \emph{Right panel:} The maximum value of the $\lambda_1(p^2,p^2,\theta)$ as a function of $\theta$.}
    \label{figlambda1max}
    \end{figure}

In Fig.~\ref{figlambda1kin} we present numerical results for $\lambda_1$, obtained from the iterative solution of  \1eq{lambdaieuc2}. The results are displayed in four panels, each for a different value of $\theta$. The diagonals of these plots 
($p_1=p_2$)
are identified with the
special kinematic configurations: 
$(\it i)$  {\it soft-gluon}, blue curve;  
$(\it ii)$ {\it totally symmetric}, 
red curve; 
$(\it iii)$ {\it quark-symmetric}, 
brown curve; and 
$(\it iv)$ {\it asymmetric}, 
yellow curve\footnote{Note that the value of $\theta$ in the totally symmetric configuration is not $2\pi/3$, since in our 
kinematics the antiquark momentum is outgoing.}.

The sequence of all diagonals, obtained as the angle $\theta$ varies within the interval 
$[0,\pi]$, may be plotted as a function of  
$\theta$, giving rise to the 
collection of curves shown 
in the left panel of 
\fig{figlambda1max}~\cite{Huber:2012kd,Cyrol:2017ewj,Blum:2017uis}.
The band is delimited by the 
soft-gluon and asymmetric 
configurations, $\theta = 0$ and 
$\theta = \pi$, respectively.

Clearly, as the angle $\theta$ increases, the peak 
of the form factors becomes more pronounced, in a continuous fashion. This pattern is 
displayed in the  
 the right panel of Fig.~\ref{figlambda1max}, 
where the four aforementioned 
kinematic configurations 
are highlighted; 
note a $27\%$ increase between the maxima of the soft-gluon and asymmetric configurations.
For large values of the momentum, 
all curves decrease logarithmically,
at the rate 
predicted in~\cite{Davydychev:2000rt}. 

Note that the main contribution to this form  factor
stems from the non-Abelian diagram $b_{\mu}$; for instance, for the soft-gluon configuration, we have  
$\lambda_1^\sg(0)=2.138$, with 
\mbox{${\mathbb B}_{1,\s R}^\sg(0)= 1.142$} and 
\mbox{${\mathbb A}_{1,\s R}^\sg(0)= -0.004$}, the rest (unity) coming from the  tree-level 
contribution.

\subsection{Non-classical form factors}

\begin{figure}[t]
    \includegraphics[scale=0.7]{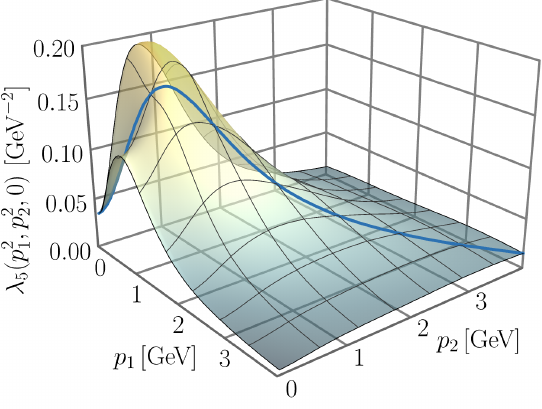} \hfil
    \includegraphics[scale=0.7]{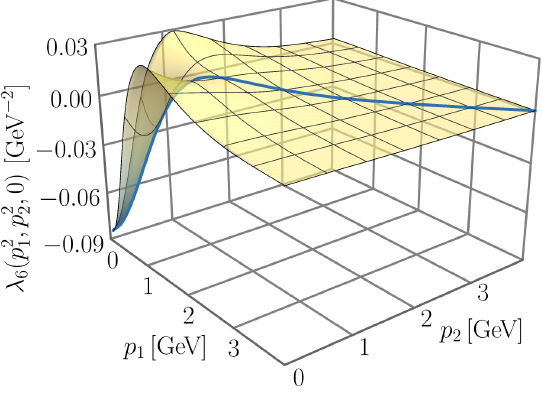}\hfil
    \includegraphics[scale=0.7]{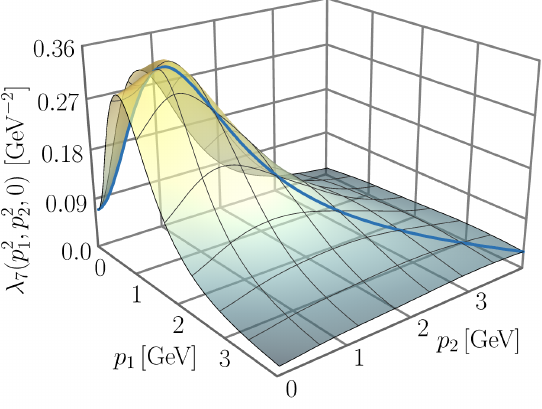} \hfil
    \caption{The chirally symmetric quark-gluon form factors $\lambda_{i}(p_1^2,p_2^{2},\theta)$, with $i= 5,6$ (upper row) and $i=7$ (lower row) plotted as functions of the magnitudes of the momenta $p_1$ and $p_2$, for a fixed value of the angle,  $\theta=0$. The blue curves 
 along the diagonals represent the corresponding soft-gluon limits of each form factor. }
    \label{3Dplots1}
\end{figure}

\begin{figure}[ht]
    \includegraphics[scale=0.7]{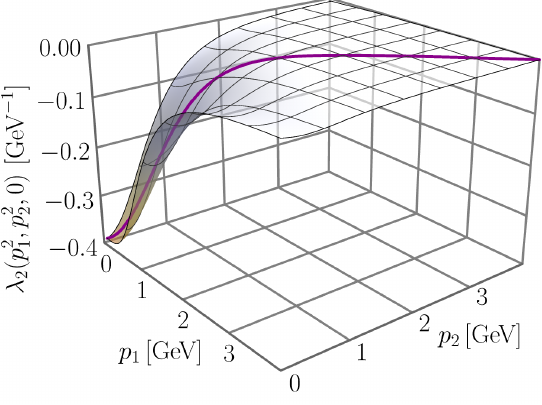} \hfil
    \includegraphics[scale=0.7]{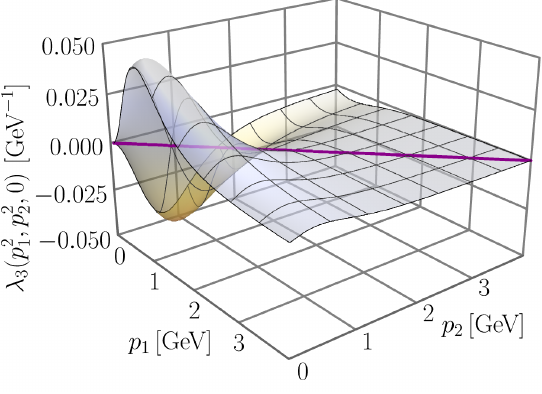} \hfil
    \includegraphics[scale=0.7]{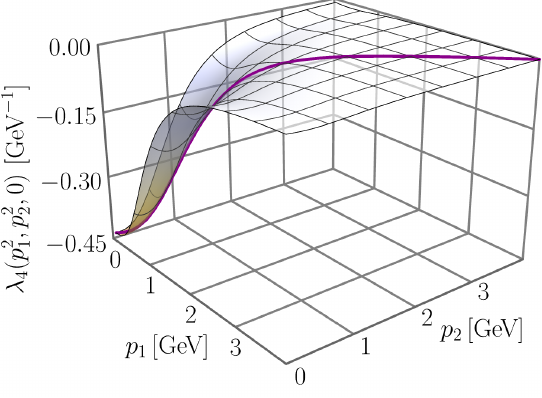} \hfil 
    \includegraphics[scale=0.7]{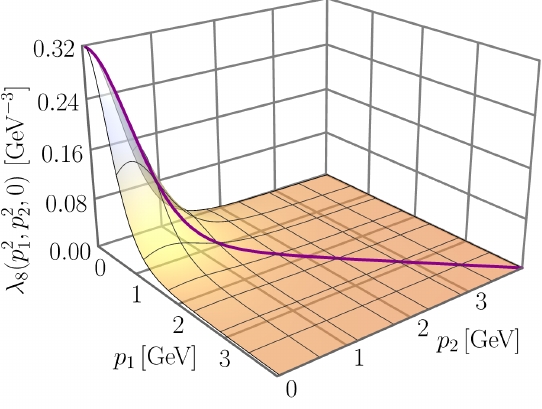} \hfil
    \caption{The chiral symmetry breaking quark-gluon form factors $\lambda_{i}(p_1^2,p_2^{2},\theta)$, with $i= 2,3$ (upper row) and $i=4,8$ (lower row) plotted as functions of the magnitudes of the momenta $p_1$ and $p_2$, for a fixed value of the angle,  $\theta=0$. The purple curves along the diagonals represent the corresponding soft-gluon limit of each form factor. }
    \label{3Dplots}
\end{figure}

Once the solution for $\lambda_1(p_1^2,p_2^2,\theta)$ in general kinematics is known, 
one may use it as input in 
\1eq{lambdaieuc2} and determine the 
remaining form factors; in fact, each 
one of them is obtained through a single integration.  
The results for the chirally symmetric 
form factors $\lambda_5$ (upper left), $\lambda_6$ (upper right), $\lambda_7$ (lower)  are shown in Fig.~\ref{3Dplots1};  while
in Fig.~\ref{3Dplots}, we present the results for the chiral symmetry breaking form factors $\lambda_2$ (upper left), $\lambda_3$ (upper right), $\lambda_4$ (lower left), $\lambda_8$ (lower right).  In each of these figures, the corresponding form factor is plotted in terms of the momenta $p_1$ and $p_2$,  fixing the value of the angle at  $\theta=0$.

%

First, notice that the form factors $\lambda_i$ with $i=4,6,7,8$ shown in  Figs.~\ref{3Dplots1} and~\ref{3Dplots} are symmetric with respect to the diagonal plane
(\mbox{$p_1 = p_2$})
(blue or purple curves, respectively). This is a
direct consequence of the charge conjugation symmetry satisfied by these form factors, as stated in Eq.~\eqref{conjC}. Observe that this property becomes visible in 3D surfaces only when $\lambda_i(p_1^2,p_2^2, \theta)$ is
plotted as a function of the momenta $p_1$ and $p_2$.  In the case of $\lambda_2$, this symmetry with respect to the diagonal is approximately satisfied, since $\lambda_3$ is very small, as shown in Fig.~\ref{3Dplots}. The fourth relation in \1eq{conjC} is also satisfied numerically.

In addition, notice that the (blue or purple) continuous curves along the diagonals in these plots represent the corresponding soft-gluon limit of each form factor, since they are defined by the condition \mbox{$p_1 = p_2$} or equivalently \mbox{$q=0$}, and the angle was fixed at $\theta=0$ in all panels.

\begin{figure}[t]
    \centering
    \includegraphics[scale=0.45]{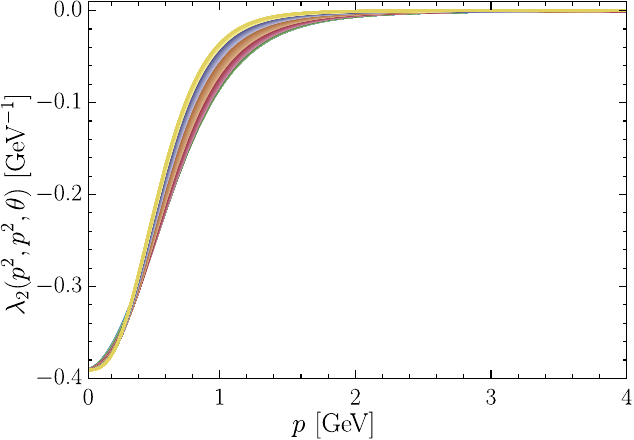}\hfil
    \includegraphics[scale=0.46]{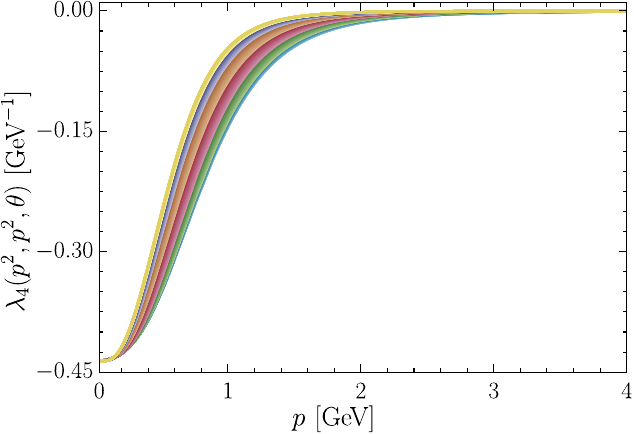}\hfil
    \includegraphics[scale=0.45]{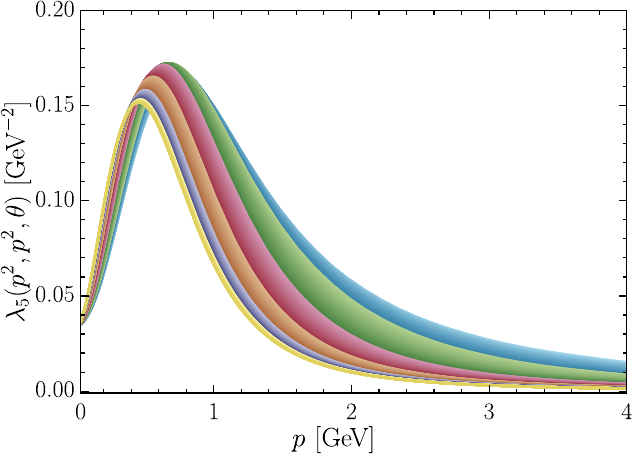}\hfil
    \vspace{0.5cm}
    \includegraphics[scale=0.47]{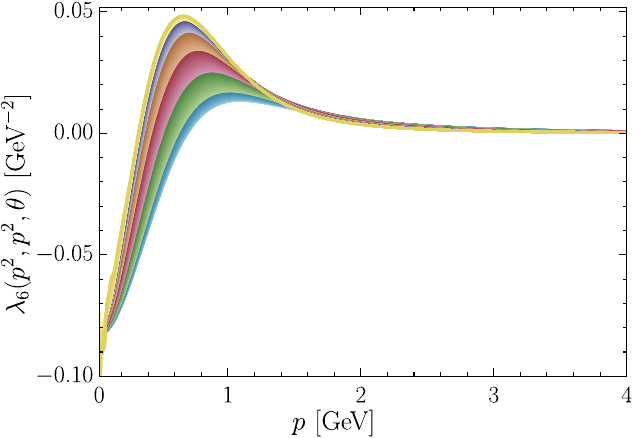}\hfil
    \includegraphics[scale=0.45]{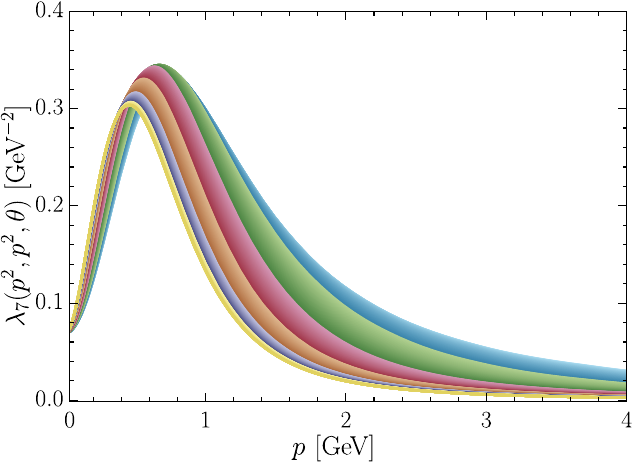}\hfil
    \includegraphics[scale=0.45]{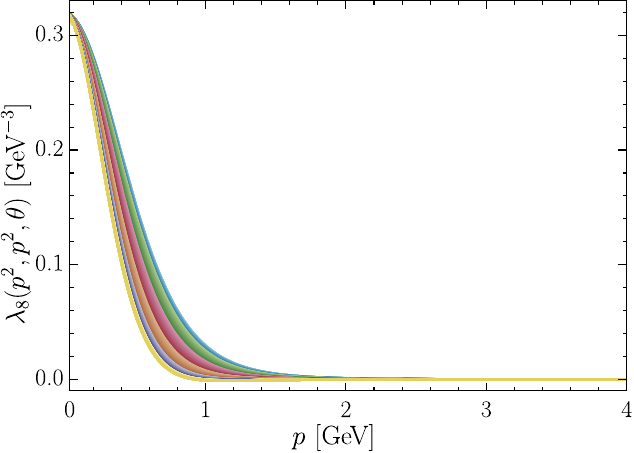}\hfil
    \vspace{0.5cm}
    \includegraphics[scale=0.7]{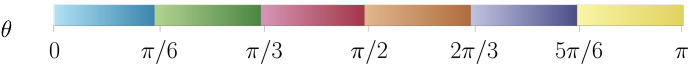}
     \caption {The angular dependence of the $\lambda_i(p_1^2,p_1^2,\theta)$ for $i=2,4,5,6,7,8$, for $p_1^2 = p_2^2 = p^2$. }
    \label{angular}
\end{figure}

In Fig.~\ref{angular}  
we display the angular dependence of all non-classical form factors when \mbox{$p_1^2=p_2^2=p^2$}, with the exception of 
$\lambda_3$, which is identically zero in this limit. As in the case of 
Fig.~\ref{figlambda1max}, all bands are 
delimited by the soft-gluon $(\theta=0)$
and asymmetric $(\theta=\pi)$ configurations.
It is evident 
 that $\lambda_2$ and $\lambda_8$ exhibit the weakest angular dependence. In contrast, 
$\lambda_4$ and $\lambda_6$  show a slightly 
stronger dependence on $\theta$, concentrated 
mainly in the momentum region where \mbox{$p\leq 2$~GeV} . Finally, $\lambda_5$ and $\lambda_7$
display a noticeably stronger dependence, practically in the entire range of momenta, although still milder than the angular dependence of $\lambda_1$
shown in Fig.~\ref{figlambda1max}. 

As a general remark we point out that 
the non-classical form factors are infrared finite and depart considerably from their 
(vanishing) tree-level values, 
approaching their expected perturbative behavior in the deep ultraviolet.

Note that  
the non-Abelian diagram dominates again 
numerically, especially 
in the cases of the form factors $\lambda_2$ and 
$\lambda_4$, where its 
contribution exceeds 
the Abelian one by 
at least one order of magnitude. 

\subsection{The quark-gluon effective couplings}

In order to 
carry out a meaningful 
comparison of 
the relative size of the 
various vertex form factors, 
it is advantageous to introduce 
dimensionless and RGI  combinations, which serve 
as generalizations of the 
traditional effective 
couplings. 
Specifically, one 
singles out special 
kinematic configurations 
(\eg soft-gluon, totally symmetric, etc), where   
the $\lambda_i$ depend on a single kinematic 
variable, and constructs  
a family of dimensionless effective couplings, 
$\widehat{g}_i(p^2)$, defined as 
\cite{Athenodorou:2016oyh,Mitter:2014wpa,Cyrol:2017ewj,Aguilar:2019uob,Gao:2021wun}{\footnote{In~\cite{Mitter:2014wpa, Gao:2021wun} the coupling $g(\mu^2)$ is included in the definition of the quark-gluon form factor $\lambda_1$, whereas here it has been factored out, see \1eq{factorG}.}
\begin{align}  
\label{lambdabar}
	 \widehat{g}_i(p^2) =  g(\mu^2)\,[p^{n_i}\lambda_i(p^2)] A^{-1}(p^2){\mathcal Z} ^{1/2}(p^2)  \,,\qquad \textrm{with}\,\,\,& n_1=0, \, \ n_{2,3,4}=1 \,, \nonumber\\
    & n_{5,6,7} = 2\,, \ n_{8} = 3\,,
\end{align}
where ${\mathcal Z}(p^2)$ and 
$A(p^2)$ are defined in Eqs.~\eqref{gluoprop} and \eqref{qprop}, respectively.

In what follows we will choose as our special configuration the soft-gluon limit, whose 
form factors are given by the diagonals of 
the 3-D plots shown in Figs.~\ref{figlambda1kin},~\ref{3Dplots1}, and~\ref{3Dplots}. 
We will therefore denote with the index 
``sg'' the corresponding form factors and effective couplings, thus implementing into \1eq{lambdabar}
the replacement
\be 
\widehat{g}_i(p^2) , \lambda_i(p^2)  \,
\longrightarrow \, \widehat{g}^{\;\sg}_i(p^2), 
\lambda^\sg_i(p^2) \,.
\label{sgrep}
\ee

In Fig.~\ref{figlambdabar} we  present the results for  $\widehat{g}^{\;\sg}_i(p^2)$, 
separating them into two subsets, those  associated with  
the $\tau_{cs}$
 (left panel),
and those related to the $\tau_{csb}$ (right panel), in accordance with the definition in  Eq.~\eqref{set_chiral}. 
It is clear from Fig.~\ref{figlambdabar} that 
the two groups of effective couplings satisfy
the hierarchies 
\be
\widehat{g}^{\;\sg}_1(p^2) > \widehat{g}^{\;\sg}_7(p^2) > \widehat{g}^{\;\sg}_5(p^2) > |\widehat{g}^{\;\sg}_6(p^2)| \,,
\qquad
|\widehat{g}^{\;\sg}_4(p^2)| > |\widehat{g}^{\;\sg}_2(p^2)|
> \widehat{g}^{\;\sg}_8(p^2) \,.
\label{hier}
\ee
Note that 
$\widehat{g}^{\;\sg}_3$ 
vanishes identically, and that $\widehat{g}^{\;\sg}_7(p^2)=2\widehat{g}^{\;\sg}_5(p^2)$, manifesting the charge conjugation symmetry of the vertex [see discussion below \1eq{conjC}].
We have verified that the hierarchies given in 
\1eq{hier} persist in the totally symmetric, quark-symmetric, and soft-gluon limits. Moreover, our results are in qualitative agreement with those presented in~\mbox{\cite{Mitter:2014wpa,Cyrol:2017ewj,Gao:2021wun}},
where the effective couplings were computed in the totally symmetric configuration.
%
\begin{figure}[t]
    \centering
    \includegraphics[scale=0.8]{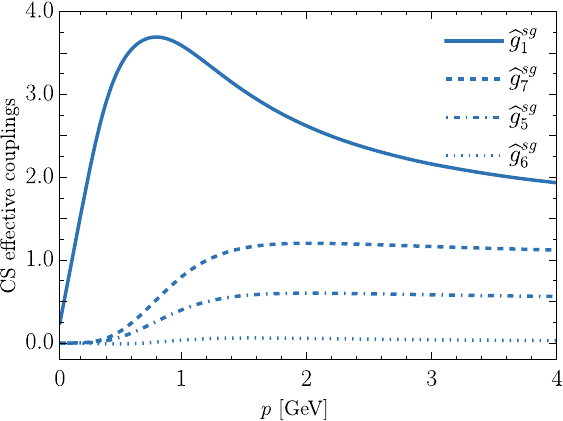}
    \hspace{0.5cm}
    \includegraphics[scale=0.84]{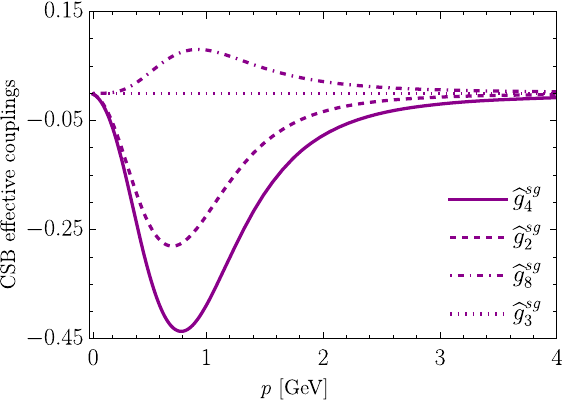}
    \caption{The quark-gluon effective couplings,  $\widehat{g}^{\;\sg}_i(p^2)$,  for the chirally symmetric form factors $\lambda_{1,5,6,7}$ (left panel), and for the chiral symmetry breaking $\lambda_{2,3,4,8}$ (right panel). The effective couplings were determined from~\1eq{lambdabar}, with the 
    $\lambda_i(p^2)$ in 
    the soft-gluon kinematics.}
    \label{figlambdabar} 
\end{figure}

If we carry out the replacement 
$\lambda_1(p^2)\rightarrow \lambda_1(p^2,p^2,\theta)$ in 
the $\widehat{g}_1(p^2)$ of 
\1eq{lambdabar}, 
 we may obtain a continuous 
family of effective charges, 
$\alpha_{qg}(p^2, \theta)$, defined as~\cite{Mitter:2014wpa,Cyrol:2017ewj}
\begin{align}
    \alpha_{qg}(p^2,\theta) = \frac{\widehat{g}_1^2(p^2,\theta)}{4\pi}\,,
\end{align}
which are shown in \fig{figalpha}. 
We note a considerable 
difference in size among the 
members of this family; 
in particular, the limiting cases 
$\theta=0$ and $\theta=\pi$ 
are separated by a factor of 3 between peaks. 

\begin{figure}[t]
    \includegraphics[scale=0.8,valign=t]{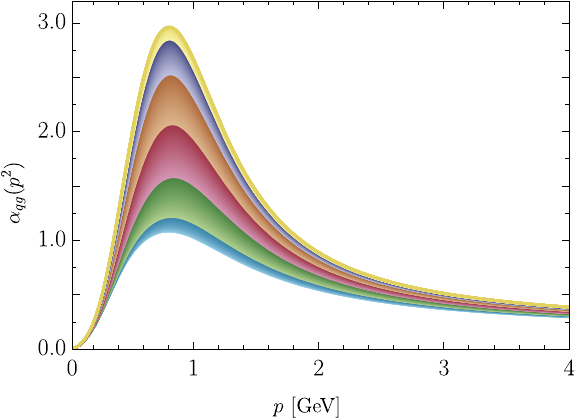}
    \hspace{1cm}
    \includegraphics[scale=0.7,valign=t]{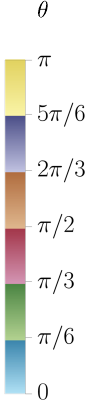}
    \caption{Family of effective charges extracted from the quark-gluon vertex, using different kinematic configurations for the classical form factor $\lambda_1(p^2,p^2,\theta)$.}
    \label{figalpha} 
\end{figure}

\subsection{Varying the inputs} 

It is particularly important to acquire a 
quantitative understanding of 
how variations of the SDE inputs 
affect the final result. 
To that end, we perform small variations around the fits to lattice data used as 
inputs,\footnote{In doing so, we keep the value 
of the strong charge fixed at 
$\alpha_s(\mu)=0.55$.} 
and consider their impact on $\lambda_1^\sg$.

Specifically, we solve \1eq{lambdaieuc2} for $\lambda_1^\sg$, each time substituting one input \mbox{$f=[\Delta, \Ls, A, \mathcal{M}]$} by $f^\pm(p^2)$, where
\begin{align}  \label{vary4}
    & \Delta^{\pm}(p^2) = \Delta(p^2) \pm \delta_1/[1+(p^2/\kappa_1^2)^2] \,, \qquad &&L^{\pm}_\sg(p^2) = L_\sg^{*}(p^2) \pm \delta_2/[1+(p^2/\kappa_2^2)^2]\,, \nonumber \\
    &1/A^{\pm}(p^2) = 1/A(p^2) \pm \delta_2/[1+(p^2/\kappa_2^2)^2] \,, \quad &&{\mathcal{M}}^{\pm}(p^2) = {\mathcal{M}}(p^2) \pm \delta_3/[1+(p^2/\kappa_2^2)^2]\,,
\end{align}
with $\delta_1=0.16{\rm \,GeV^{-2}}$, 
$\delta_2=0.03$, 
and $\delta_3=0.015 {\rm \,GeV}$, $\kappa_1^2=0.5{\rm \,GeV^2}$, and $\kappa_2^2=2{\rm \,GeV^2}$. 
With this choice of parameters, 
the infrared finite ingredients, $\Delta^\pm$, $1/A^\pm$, and $\mathcal{M}^\pm$ differ from their central values
by $\sim4\%$, within the 
interval $p\in[0,1]$~GeV, while  
$\Ls^\pm(p^2)$, which diverges at the 
origin, shows the same  
variation for $p\in[0.1,1]$~GeV. 
All $f^\pm(p^2)$ approaches $f(p^2)$ rapidly for $p\gtrapprox 2$~GeV.

The $\lambda_1^\sg$ resulting from performing each of the above variations is shown as a separate panel in \fig{bands}; the inset in each case displays the ingredient being varied. In order to visually track the direction of the variations in $\lambda_1^\sg$, the lower bound, $f^-$, of any given ingredient,  and the $\lambda_1$ resulting from using this $f^-$, are marked with a dashed line. We note that $\lambda_1^\sg$ is enhanced when increasing $\Delta$, $\Ls$, or the quark wave function, $1/A$. On the other hand, as expected, an increase in $\mathcal{M}$ reduces $\lambda_1^\sg$.

Since the numerically dominant
non-Abelian diagram (see~Sec.~\ref{sec:classical})
is linear in $1/A$ and $\Ls$, but quadratic in $\Delta$, we expect 
$\lambda_1^\sg$
to be especially sensitive to the gluon dressing function.
Indeed, we observe in \fig{bands} that a variation of $4\%$ in $1/A$ ($\Ls$) in the infrared region, has a $5\%$ ($3\%$) effect on $\lambda_1^\sg(0)$. Moreover, $\lambda_1^\sg$ is rather insensitive to small variations of the constituent quark mass, $\mathcal{M}$, changing by only $1\%$ in our tests. In contrast, enhancing $\Delta$ by the same $4\%$ increases $\lambda_1^\sg$ by $14\%$.

Note finally that the above 
variations are uncorrelated, 
in contradistinction to what 
happens  
in a self-contained SDE analysis, 
where a change to one of them affects 
all others in complicated ways; 
nonetheless, we hope that the 
main tendencies are correctly captured.

%
\begin{figure}[t]
    \centering
    \setbox1=\hbox{\includegraphics[height=5cm]{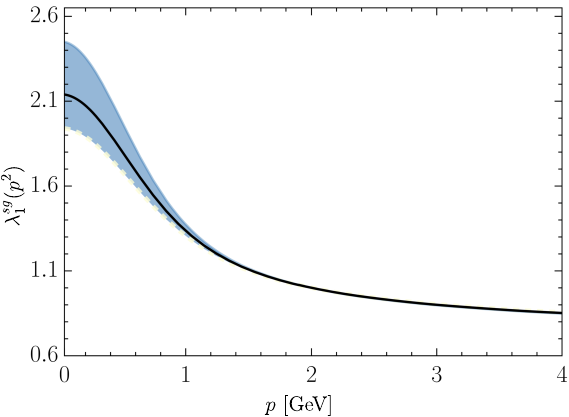} \hfil}
    \includegraphics[scale=0.7]{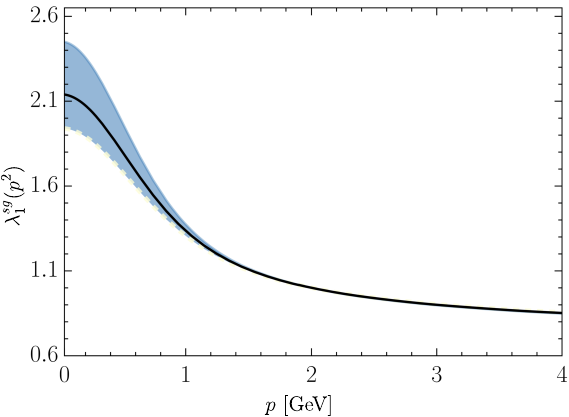}
    \hspace{-0.5cm}\llap{\makebox[\wd1][r]{\raisebox{1.9cm}{\includegraphics[scale=0.37]{fig12a}}}}\hfil
    %
    \setbox1=\hbox{\includegraphics[height=5cm]{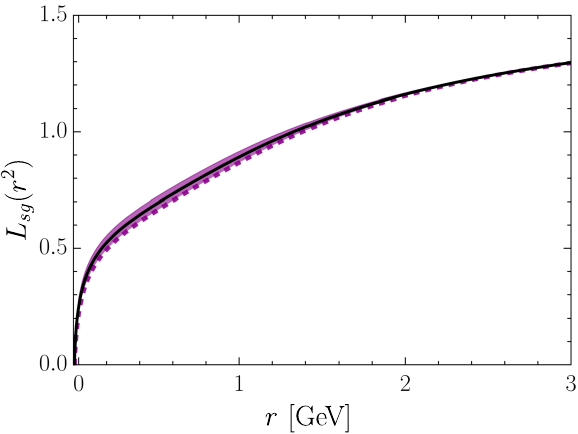} \hfil}
    \includegraphics[scale=0.7]{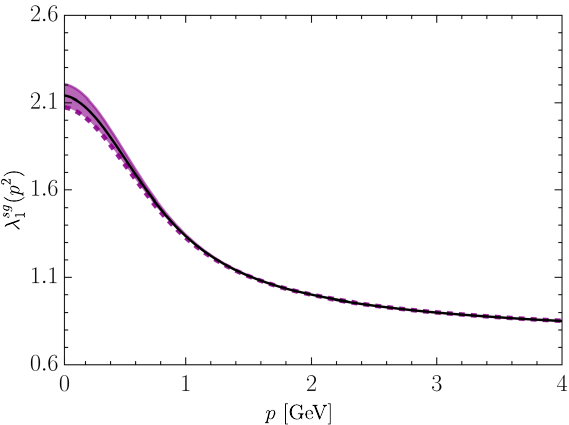}
    \hspace{-0.5cm}\llap{\makebox[\wd1][r]{\raisebox{1.9cm}{\includegraphics[scale=0.37]{fig12c}}}}
    %
    \setbox1=\hbox{\includegraphics[height=5cm]{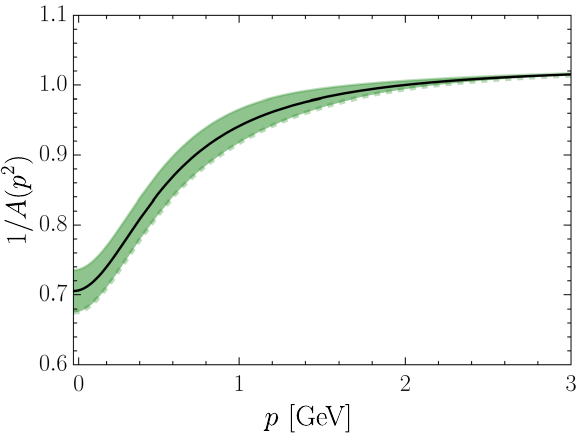} \hfil}
    \includegraphics[scale=0.7]{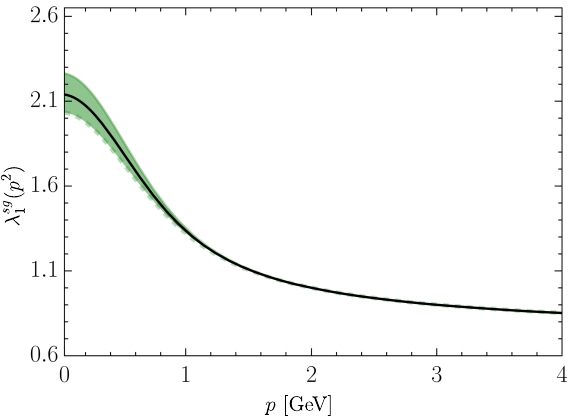}
    \hspace{-0.5cm}\llap{\makebox[\wd1][r]{\raisebox{1.9cm}{\includegraphics[scale=0.37]{fig12f}}}}\hfil
    %
    \setbox1=\hbox{\includegraphics[height=5cm]{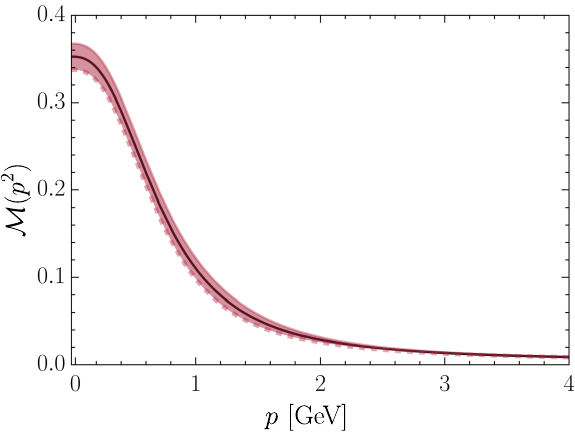} \hfil}
    \includegraphics[scale=0.7]{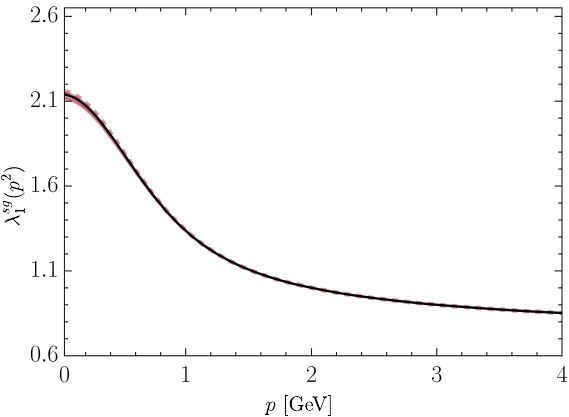}
    \hspace{-0.5cm}\llap{\makebox[\wd1][r]{\raisebox{1.9cm}{\includegraphics[scale=0.37]{fig12h}}}}
    \caption{The effect of varying the inputs 
    ${\cal Z}(q^2)$ (upper left), $\Ls(r^2)$ (upper right), 
    $1/A(p^2)$ (lower left), and $\mathcal{M}(p^2)$ (lower right).  
    The insets depict the corresponding variations, while 
    the main plots show their impact on $\lambda_1$. }
    \label{bands}
\end{figure}

\subsection{Comparison with the lattice} 
\label{sec:comp_lattice}

In this subsection we compare our results for the quark-gluon vertex in the soft-gluon configuration with those obtained in the $N_f=2$ lattice simulation of~\cite{Kizilersu:2021jen}.

We first recall that in the soft-gluon limit, four out of the eight tensor structures given by~\eqref{Taus} vanish identically. Upon setting $p_1=p_2=p$, the tensorial structure of the transversely-projected vertex, $\overline{\fatg}_\mu(0,p,-p)$, simplifies to Eq.~\eqref{softgluon1}, 
where the vanishing of  $\lambda^\sg_3(p^2)$ 
has already been taken into account. 
From the analysis presented
in the Appendix~\ref{MtoE}, 
it is clear that, 
formally, the classical 
form factors of the SDE and the lattice are  exactly the same, \ie  \mbox{$\lambda^\sg_1(p^2)=\lambda_1^{\rm L}(p^2)$}, as dictated by 
\1eq{lattice1}.

\begin{figure} [t]
    \centering
    \includegraphics[scale=0.8]{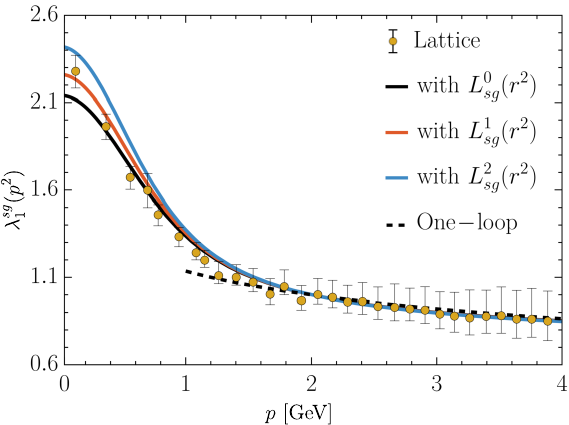}\hfill
    \includegraphics[scale=0.8]{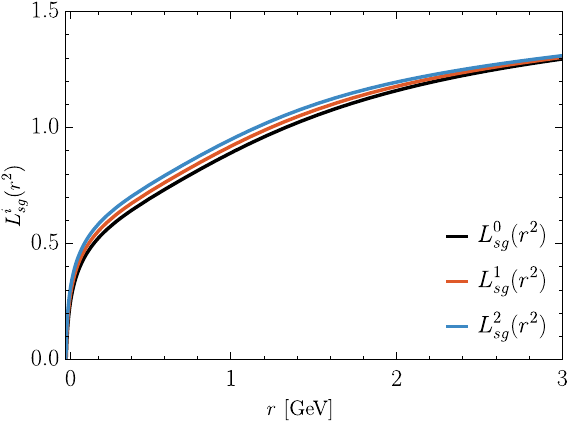}
    \caption{{\it Left panel:} Comparison of the form factor $\lambda_1(p^2)$ in the soft-gluon configuration (continuous curves), computed with the $\Ls^i(r^2)$ shown in the right panel, with the ``L08'' lattice data  (circles) of~\cite{Kizilersu:2021jen}. The one-loop result of~\cite{Davydychev:2000rt} is shown as a black dashed line. {\it Right panel:} The $\Ls^i(r^2)$ given by \1eq{eqbands} used to compute $\lambda_1^\sg(p^2)$. 
    }
     \label{figlattice} 
\end{figure}

In the left panel of Fig.~\ref{figlattice}, 
we compare the  
SDE result for the form factor $\lambda^\sg_1(p^2)$, 
(black continuous curves) with the ``L08'' lattice data of~\cite{Kizilersu:2021jen} (circles).  
Evidently, 
both curves nearly coincide over most of the momentum range, showing a minor departure only in 
the deep infrared, 
where the corresponding saturation points differ by approximately $7\%$.
Note that the value used in the SDE for obtaining this solution is 
$\alpha_s(\mu^2) = 0.55$.  
Finally, 
the black dashed line 
represents the one-loop result for $\lambda^\sg_1(p^2)$ taken 
from~\cite{Davydychev:2000rt}, duly renormalized in the \MOMt{} scheme; 
 we observe that both the SDE and lattice results recover the perturbative behavior for $p \gtrapprox 2$~GeV. }

Given the analysis of the previous subsection, it is clear that the
coincidence between SDE and lattice  
may be easily improved even further, 
by introducing slight  deviations in 
the inputs employed in the SDE. 
To demonstrate this possibility with 
a concrete example, let 
us allow for  minor variations 
of the $\Ls(r^2)$ around 
$\Ls^{*}(r^2)$, using the same functional dependence as in 
\1eq{vary4}.

In particular, we 
consider the family $\Ls^{i}(r^2)$, given by 
\begin{align}   
\label{eqbands}
   L^{i}_\sg(r^2) = L^{*}_\sg(r^2)  \,+ \,
\frac{\epsilon_{i}}{1+(r^2/\kappa^2)^2}   
\,,  \qquad {i=0,1,2\,,}
\end{align}
with $\epsilon_0=0$, $\epsilon_1=0.03$, $\epsilon_2=0.06$, and, $\kappa^2=5{\rm \,GeV^2}$; the three curves are 
represented together in the right panel of Fig.~\ref{figlattice}.
The resulting $L^{1}_\sg(r^2)$ 
and $L^{2}_\sg(r^2)$ clearly improve 
the overall coincidence  with the lattice in the deep infrared. 
Similar 
levels of agreement may be 
obtained by varying the other inputs,
in the spirit of \1eq{vary4}; we 
will not pursue this possibility 
any further.

%
\begin{figure}[t]
    \centering
    \includegraphics[width=0.45\linewidth]{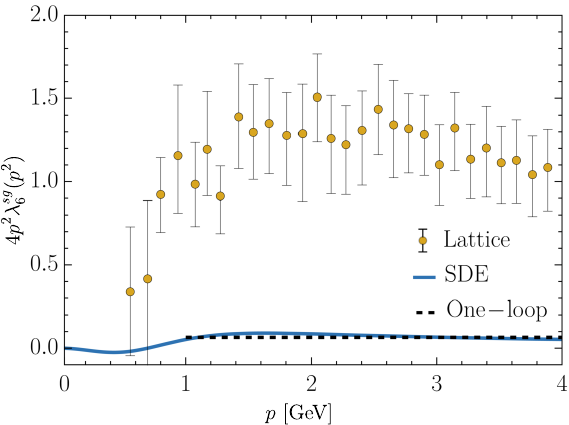}\hfil
    \includegraphics[width=0.465\linewidth]{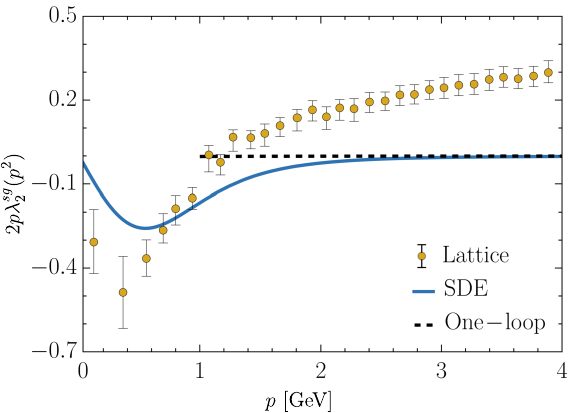}
    \caption{Comparison of the dimensionless soft-gluon form factors $4p^2\lambda_6^\sg(p^2)$ (left) and  $2p\lambda_2^\sg(p^2)$ (right) (blue continuous curves) with lattice data of~\cite{Kizilersu:2021jen} (circles). We also plot the corresponding one-loop results from~\cite{Davydychev:2000rt} (black dashed curve).}
    \label{figlattice2}
\end{figure}

The excellent agreement
found in the case of 
$\lambda^\sg_1(p^2)$
between the SDE results and the lattice 
is in stark contrast to what happens with the 
remaining two form factors of the soft-gluon configuration. 
Specifically, using the formal correspondence 
\mbox{$\lambda^\sg_6(p^2) = \lambda_2^{\rm L}(p^2)$} and  
\mbox{$\lambda^\sg_2 (p^2)=
\lambda_3^{\rm L}(p^2)$} 
[see \1eq{softgluoneuc}],
in \fig{figlattice2}
we compare 
the typical 
dimensionless combinations 
$4p^2 \lambda^\sg_6(p^2)$ and 
$2 p \lambda^\sg_2(p^2)$,  
observing strong qualitative discrepancies with respect to the lattice data of~\cite{Kizilersu:2021jen}. 

Historically, 
the lattice determination of 
the form factor $4p^2\lambda^\sg_6(p^2)$ [left panel 
of \fig{figlattice2}]
has been rather problematic. In the 
quenched simulations of~\cite{Skullerud:2003qu}, $4p^2\lambda^\sg_6(p^2)$ appears to diverge at the origin, 
contrary to all continuous studies 
that yield a vanishing result~\mbox{\cite{Bhagwat:2004kj,Llanes-Estrada:2004hnb,Aguilar:2014lha,Williams:2015cvx,Aguilar:2016lbe}}.
 Although the $N_f=2$ simulation appears 
 to have ameliorated this flaw, 
 the ultraviolet tail of the data is clearly at odds with the correct one-loop perturbative behavior~\cite{Davydychev:2000rt}, shown as black dashed in \fig{figlattice2}. 
It would seem, therefore, that 
additional analysis is required before
a meaningful comparison with the SDE results 
may be conducted. 

Regarding $2p\lambda_2^\sg(p^2)$,  shown in the right panel of Fig.~\ref{figlattice2}, we again observe a discrepancy between the lattice data and the expected one-loop behavior~\cite{Davydychev:2000rt}. Although our result for $2p\lambda_2^\sg(p^2)$ exhibits a qualitative pattern similar to the unquenched lattice data~\cite{Kizilersu:2021jen}, 
the minimum found in the deep infrared is considerably shallower than the one observed on the lattice.

It is worth stressing that the lattice $\lambda_2^\sg(p^2)$ and $\lambda_6^\sg(p^2)$ are known to be more severely affected by discretization artifacts than $\lambda_1^\sg(p^2)$. Indeed, the ``tree-level correction'' procedure employed in~\cite{Kizilersu:2021jen},
for the purpose of reducing discretization artifacts, improves significantly the agreement of $\lambda_1^\sg(p^2)$ with its perturbative behavior. However, as explicitly stated in~\cite{Kizilersu:2021jen}, it has the opposite effect on $\lambda_2^\sg(p^2)$, pushing it away from its one-loop result at large momenta. Moreover, while for $\lambda_1^\sg(p^2)$ and $\lambda_2^\sg(p^2)$ the discretization artifacts are apparent mostly in the ultraviolet (see Figs.~4 and 6 of~\cite{Kizilersu:2021jen}), the aforementioned procedure affects $\lambda_6^\sg(p^2)$ within the entire range of momentum (see Fig.~5 of~\cite{Kizilersu:2021jen}).

\section{Verifying multiplicative renormalizability}
\label{multren} 

In this section we 
probe the veracity 
of multiplicative 
renormalizability 
at the level of the 
form factor $\lambda^\sg_1(p^2)$.
This exercise is particularly 
relevant in view of the 
subtractive nature of 
the renormalization procedure
employed on the SDE 
derived from the 
3PI effective action formalism, concretely 
Eqs.~\eqref{lambdaieuc2} and \eqref{Z1final}, 
as discussed 
in Sec.~\ref{sderen}. 
The upshot of these considerations is that 
multiplicative renormalizability is 
faithfully reflected in 
the solutions of 
the SDE, essentially 
due to the fact that 
all vertices entering in 
diagrams $a_{\mu}$ and $b_{\mu}$ of  \fig{fig:sde}
are fully dressed.

The main idea of our procedure is to 
repeat the calculation of $\lambda_1$, with \1eq{lambdaieuc2} renormalized in the same scheme, \MOMt{}, but at a different renormalization point, to be denoted by $\nu$, \ie imposing 
\2eqs{ren_conds}{Z1final}
with $\mu\to\nu$. 
Then, if multiplicative 
renormalizability 
is respected, 
the two answers must be 
related by 
\be  
\label{lambdamunu}
\lambda^\sg_1(p^2,\mu^2) = 
    \frac{\lambda^\sg_1(p^2,\nu^2)}{\lambda^\sg_1(\mu^2,\nu^2)} \,.
\ee
In order to proceed, we need the inputs for $\Delta(q^2)$, $\Ls(r^2)$, and $A(p^2)$, 
shown in Figs.~\ref{inputsDL} and~\ref{inputsAM}, respectively, at different
renormalization points~\footnote{Notice that the quark dynamical mass, ${\mathcal M}(p^2)$,  is a $\mu$-independent quantity, and therefore does not need to be rescaled.}. 
To obtain them, we 
will assume that multiplicative renormalizability is valid for these three functions,  
and use relations analogous to  
\1eq{lambdamunu} to deduce 
their form at the new renormalization point. 

Specifically, it follows from \1eq{renconst}, and the fact that the unrenormalized Green's functions do {\it not} depend on 
the renormalization point, that
\begin{align}
    \Delta(q^2,\nu^2) =&\, \Delta(q^2,\mu^2)\frac{Z_A(\mu^2)}{Z_A(\nu^2)}\,, \quad  &A(p^2,\nu^2) =&\, A(p^2,\mu^2)\frac{Z_F(\nu^2)}{Z_F(\mu^2)}\,, \nonumber\\
    \lambda_1^\sg(p^2,\nu^2) =&\, \lambda_1^\sg(p^2,\mu^2) \frac{Z_1(\nu^2)}{Z_1(\mu^2)}\,, \quad  &\Ls(s^2,\nu^2) =&\, \Ls(s^2,\mu^2) \frac{Z_3(\nu^2)}{Z_3(\mu^2)}\,. \label{rescale_gen}
\end{align}
Now, the values $\Delta(\nu^2,\nu^2)$, $A(\nu^2,\nu^2)$, and $\lambda_1^\sg(\nu^2,\nu^2)$, are fixed by the renormalization prescription of \1eq{ren_conds} with $\mu \to \nu$. Hence, evaluating \1eq{rescale_gen} at $q^2 = p^2 = \nu^2$ entails
\be 
\frac{Z_A(\mu^2)}{Z_A(\nu^2)} = \frac{1}{\nu^2\Delta(\nu^2,\mu^2)} \,, \qquad \frac{Z_F(\nu^2)}{Z_F(\mu^2)} = \frac{1}{A(\nu^2,\mu^2)} \,, \qquad \frac{Z_1(\nu^2)}{Z_1(\mu^2)} = \frac{1}{\lambda_1^\sg(\nu^2,\mu^2)} \,. \label{Zs_ratios}
\ee
Then, substituting into \1eq{rescale_gen} we obtain \1eq{lambdamunu} with $\mu\leftrightarrow \nu$, together with 
\begin{align}
    \Delta(q^2,\nu^2) = \frac{\Delta(q^2,\mu^2)}{\nu^2 \Delta(\nu^2,\mu^2)}\,, \quad 
    \quad A(p^2,\nu^2) = \frac{A(p^2,\mu^2)}{A(\nu^2,\mu^2)}\,. \label{rescale_Delta_A}
\end{align}
On the other hand, the renormalization conditions of \1eq{ren_conds} do not specify the value of $\Ls(\nu^2,\nu^2)$. Instead, to determine $Z_3(\nu^2)/Z_3(\mu^2)$, we employ the fundamental relation 
$Z_3 = Z_1 Z_A/Z_F$, derived directly 
from \1eq{eq:sti_renorm}, 
together with \1eq{Zs_ratios}, which yield
\be 
\frac{Z_3(\nu^2)}{Z_3(\mu^2)} = \frac{\nu^2\Delta(\nu^2,\mu^2)A(\nu^2,\mu^2)}{\lambda_1^\sg(\nu^2,\mu^2)} \,,
\ee
and so, 
\be 
\Ls(s^2,\nu^2) = \frac{\nu^2\Delta(\nu^2,\mu^2)A(\nu^2,\mu^2)}{\lambda_1^\sg(\nu^2,\mu^2)} \Ls(s^2,\mu^2)\,. \label{rescale_L}
\ee

The next step is to relate the values of $\alpha_s(\nu^2)$ and $\alpha_s(\mu^2)$, which we achieve by invoking the effective coupling $\widehat{g}^{\;\sg}_1(p^2)$ of \1eq{lambdabar}. Since $\widehat{g}^{\;\sg}_1(p^2)$ is RGI, its value is the same when computed with ingredients renormalized at either $\mu$ or $\nu$, \ie
\be 
\widehat{g}^{\;\sg}_1(p^2) = \frac{g(\mu^2)\lambda^\sg_1(p^2,\mu^2){\mathcal Z}^{1/2}(p^2,\mu^2)}{A(p^2,\mu^2)}= \frac{g(\nu^2)\lambda^\sg_1(p^2,\nu^2){\mathcal Z}^{1/2}(p^2,\nu^2)}{A(p^2,\nu^2)} \,.
\ee
Therefore, setting $p = \nu$ in the above, using the renormalization prescription of \1eq{ren_conds} with $\mu\to \nu$, and $g^2 = 4\pi\alpha_s$, leads to
\be 
\alpha_s(\nu^2) = \alpha_s(\mu^2)\left[ \lambda_1^\sg(\nu^2,\mu^2) \right]^2 A^{-2}(\nu^2,\mu^2) \nu^2\Delta(\nu^2,\mu^2) \,. \label{rescale_alpha}
\ee

Then, using the previously obtained curve for $\lambda_1^\sg(p^2,\mu^2)$, shown in \fig{figlattice}, together with the external inputs renormalized at $\mu = 2$~GeV, discussed in items {\it (i-iv)} of Subsec.~\ref{inp}, we get all required inputs renormalized at $\nu = 4.3$~GeV through \3eqs{rescale_Delta_A}{rescale_L}{rescale_alpha}. In particular, we find $\alpha_s(\nu^2) = 0.28$.

\begin{figure}[ht]
    \includegraphics[scale=0.8]{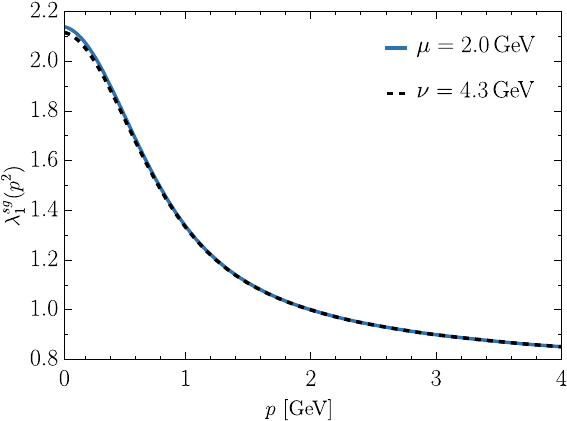} \hfil
    \caption{Verification that our solution for $\lambda_1^\sg(p^2)$ satisfies the multiplicative renormalizability property given by Eq.~\eqref{lambdamunu}.}
    \label{VarLsg}
\end{figure}

With all the necessary inputs in hand, we solve Eqs.~\eqref{lambdaieuc2} and \eqref{Z1final} again to obtain $\lambda_1(p^2,\nu^2)$ at $\nu = 4.3$~GeV. The result is then rescaled to $\mu = 2$~GeV using \1eq{lambdamunu}, and compared to the $\lambda_1(p^2,\mu^2)$ obtained by solving the SDE directly at $\mu$. The comparison is shown in \fig{VarLsg}, where the blue continuous curve shows the result of the SDE renormalized at $\mu$, whereas the black dashed corresponds to the solutions of the SDE renormalized at $\nu$ and subsequently rescaled to $\mu$ through \1eq{lambdamunu}. The nearly perfect coincidence of the two curves confirms that indeed multiplicative renormalizability is satisfied in our truncation.

\section{Conclusions} 
\label{conclusions}

In the present work we have 
studied 
the transversely-projected quark-gluon vertex of QCD with two light degenerate quarks, in the Landau gauge.
Our analysis is based on the 
SDE derived within the 3PI formalism at the three-loop level, 
where lattice results 
have been employed for 
all SDE components, 
except the quark-gluon vertex itself.

Under certain simplifying assumptions, 
we obtain
all eight form factors,
$\lambda_i$,
of this vertex 
for arbitrary space-like momenta. 
Importantly, the classical form factor, $\lambda_1$, exhibits a considerable angular dependence, displaying a large peak at the ``asymmetric limit'', which is absent in the ``soft-gluon'' configuration (see \fig{figlambda1max}). Moreover, 
our results confirm the hierarchy of the RGI effective couplings $\widehat{g}_i(p^2)$ found in 
the functional renormalization group
analysis 
of~\cite{Mitter:2014wpa,Cyrol:2017ewj},
and the SDE study of 
\cite{Gao:2021wun}. 
In particular, $\widehat{g}_1^{\;\sg}(p^2)$ is clearly dominant, its maximum value 
exceeding that of the 
second-largest $\widehat{g}_7^{\;\sg}(p^2)$
by a factor of 3. In addition, 
in the soft-gluon limit, 
the comparison with the lattice results 
of~\cite{Kizilersu:2021jen}
reveals excellent agreement 
for the classical form factor $\lambda_1^\sg(p^2)$, 
and rather strong discrepancies for 
$\lambda_2^\sg(p^2)$ and  $\lambda_6^\sg(p^2)$.

The aforementioned strong angular dependence of $\lambda_1$
precludes the possibility of accurately describing $\lambda_1$ in terms of a single variable, 
\eg \, $s^2 = (q^2+p^2_1+p^2_2)/2$, 
in contradistinction to the approximate 
``planar degeneracy'' displayed by 
the form factors of the vertices 
with three (see Sec.~\ref{simp}) and  
four-gluons~\cite{Aguilar:2024fen,Aguilar:2024dlv}. 
In hindsight, this difference appears 
natural, given that the planar degeneracy hinges crucially on the Bose symmetry of the vertex in question, which is clearly absent in the case of the quark-gluon vertex. To be sure, the analysis of~\cite{Aguilar:2023mam}, carried out under the hypothesis of planar degeneracy for the quark-gluon vertex, needs to be revisited, and the full kinematic dependence of $\lambda_1(q,p_2,-p_1)$ obtained here must be properly taken into account.

The numerical treatment in this work has been 
greatly simplified by decoupling the 
system of eight integral 
equations given in \1eq{lambdaieuc2}; 
in particular, 
the full dependence  
of all SDE kernels on $\lambda_1$ 
was maintained, while all other 
form factors were dropped. 
An evident improvement 
on the present  analysis may be 
achieved by restoring 
the full dependence of 
the $\lambda_i$, and solving the 
resulting system of equations;
calculations in this direction are already in 
progress.

 It is especially challenging  
to couple the quark-gluon vertex 
to the gap equation of the 
quark propagator, 
and solve the resulting system 
self-consistently, maintaining the full 
kinematic structure of the $\lambda_i$. 
Evidently, such a treatment 
would determine dynamically 
the functions 
$A(p^2)$ and ${\mathcal M}(p^2)$, 
eliminating the need of 
using lattice inputs for them, as was done here. 
 Within such a framework, one may 
explore the impact of the  
$\lambda_i$ found in this work 
on the standard parameters 
describing dynamical chiral symmetry 
breaking, such as the 
constituent quark mass, 
the pion decay constant, and the 
light chiral condensate, 
see~\cite{Gao:2021wun} and references 
therein.  In fact, it would be 
interesting to study 
the effect 
that the pronounced 
angular dependence of 
$\lambda_1$  might have on chiral 
dynamics.
We hope to be able to address some of 
these open issues in the near future.

\section{Acknowledgments}
\label{sec:acknowledgments}

We thank Fei Gao 
and Jan Pawlowski for several useful interactions.
The work of  A.~C.~A. and  G. L. T. is supported by the CNPq grants \mbox{310763/2023-1} and \mbox{131385/2022-4}.
A.~C.~A also acknowledges financial support from project 464898/2014-5 (INCT-FNA). The work of B. M. O. was financed in part by the Coordenação de Aperfeiçoamento de Pessoal de Nível Superior - Brasil (CAPES) - Finance Code 00, and by the CNPq grants \mbox{141409/2021-5}.  M.~N.~F. acknowledges financial support from the National Natural Science Foundation of China (grant no. 12135007). J.~P. is supported by the Spanish MICINN grant PID2020-113334GB-I00  and the Generalitat Valenciana grant CIPROM/2022/66. This research was performed using the Feynman Cluster of the
John David Rogers Computation Center (CCJDR) in the Institute of Physics Gleb
Wataghin, University of Campinas.

\appendix

\section{Transformation rules from 
Minkowski to Euclidean space} 
\label{MtoE}

In this Appendix, we derive 
the relations 
between the form factors of the 
Minkowski tensor basis of \2eqs{decomp}{Taus} and those of the 
direct Euclidean 
decompositions given by~\cite{Mitter:2014wpa,Kizilersu:2021jen}. 

As mentioned in the main text, the basis of \1eq{Taus} ensures, by construction, the equivalence between our form factors and those of~\cite{Mitter:2014wpa}, upon passing to Euclidean space. 
In~\cite{Mitter:2014wpa}, the 
Euclidean vertex, $\overline{\fatg}^\euc_\mu(q^\euc,p^\euc_2,-p^\euc_1)$, is decomposed directly as
\begin{align} \label{decomp2}
    \overline{\fatg}^\euc_\mu(q^\euc,p^\euc_2,-p^\euc_1)=\sum_{i=1}^{8}\lambda^\euc_i(q^\euc,p^\euc_2,-p^\euc_1)P_{\mu\nu}(q^\euc)\tau_{i\,\euc}^\nu(p^\euc_2,-p^\euc_1) \,, 
\end{align}
with
\begin{align} 
    &\tau^\nu_{1\,\euc}(p^\euc_2,-p^\euc_1) =\gamma_\euc^\nu\,, \quad &&\tau^\nu_{2\,\euc}(p^\euc_2,-p^\euc_1) = i (p^\euc_1+p^\euc_2)^\nu\,, \nonumber \\
    &\tau^\nu_{3\,\euc}(p^\euc_2,-p^\euc_1) =i (\slashed{p}^\euc_1+\slashed{p}^\euc_2)\gamma_\euc^\nu\,, \quad &&\tau^\nu_{4\,\euc}(p^\euc_2,-p^\euc_1) = i(\slashed{p}^\euc_2-\slashed{p}^\euc_1)\gamma_\euc^\nu\,,\nonumber\\
    &\tau^\nu_{5\,\euc}(p^\euc_2,-p^\euc_1) =  (\slashed{p}^\euc_1-\slashed{p}^\euc_2)(p^\euc_1+p^\euc_2)^\nu\,, \quad &&\tau^\nu_{6\,\euc}(p^\euc_2,-p^\euc_1) =-(\slashed{p}^\euc_1+\slashed{p}^\euc_2)(p^\euc_1+p^\euc_2)^\nu\,,\nonumber\\
    &\tau^\nu_{7\,\euc}(p^\euc_2,-p^\euc_1) = -\frac{1}{2}[\slashed{p}^\euc_1,\slashed{p}^\euc_2]\gamma_\euc^\nu\,, \quad &&\tau^\nu_{8\,\euc}(p^\euc_2,-p^\euc_1) = -\frac{i}{2}[\slashed{p}^\euc_1,\slashed{p}^\euc_2](p^\euc_1+p^\euc_2)^\nu \,.
\label{mittbas}    
\end{align}

To demonstrate that the basis above is the Euclidean equivalent of \2eqs{decomp}{Taus}, we follow the procedure described in~\cite{Skullerud:2002ge}. 

We begin by contracting \1eq{decomp} from the right\footnote{Since $\gamma_\mu\tau^\mu_{7} = 0$, contracting with $\gamma_\mu$ from the left leaves the Euclidean form of $\tau^\mu_{7}$ undetermined.} by the Minkowski Dirac matrix, $\gamma^\mu$, and transforming the result to Euclidean space, to obtain $\left[\overline{\fatg}_\mu(q,p_2,-p_1)\gamma^\mu\right]_\wic$. Then, we verify that the result is the same as contracting \1eq{decomp2} by the Euclidian Dirac matrix, $\gamma_\euc^\mu$, \ie
\begin{align} \label{Gammag}
    \left[\overline{\fatg}_\mu(q,p_2,-p_1)\gamma^\mu\right]_\wic=\overline{\fatg}^\euc_\mu(q,p_2,-p_1)\gamma_\euc^\mu\,.
\end{align}

For simplicity, we illustrate this procedure below by retaining only the contributions from the form factors $\lambda_1$ and $\lambda_3$; generalizing to the remaining $\lambda_i$ is straightforward. 

Contracting \1eq{decomp} from the right with $\gamma^\mu$ one obtains
\begin{align}
    \overline{\fatg}_\mu(q,p_2,-p_1)\gamma^\mu = (d-1)\left[ \lambda_1(q,p_2,-p_1)+ \lambda_3(q,p_2,-p_1)(\slashed{p}_1+\slashed{p}_2) \right]+ \ldots\,,
\end{align}
where $d$ is the spacetime dimension. Using the standard rules to convert the above result to Euclidean space~\cite{Roberts:1994dr}, and requiring that the form factors do not change sign in the process, \ie
\begin{align}
    \slashed{p} \rightarrow i \slashed{p}_\euc\,, \qquad p^2 \rightarrow -p^2_\euc\,, \qquad  \lambda_i(q,p_2,-p_1) \rightarrow \lambda^\euc_i(q^\euc,p^\euc_2,-p^\euc_1)\,, \label{euc_transf}
\end{align}
one gets that 
\begin{align}
    \left[\overline{\fatg}_\mu(q,p_2,-p_1)\gamma^\mu\right]_\wic = (d-1)\left[ \lambda^\euc_1(q^\euc,p^\euc_2,-p^\euc_1)-i \lambda^\euc_3(q^\euc,p^\euc_2,-p^\euc_1)(\slashed{p}^\euc_1+\slashed{p}^\euc_2) \right]+ \ldots\,.
\end{align}

An additional step may be required,  depending on the conventions adopted for the Euclidean quark propagator. Applying \1eq{euc_transf} to \1eq{S0}, we obtain $S^{-1}_{0\euc}(p_\euc) = i\slashed{p}_\euc-m_q$. However, \cite{Mitter:2014wpa} adopts the convention that $S^{-1}_{0\euc}(p_\euc) = i \slashed{p}_\euc+m_q$. These two conventions are related by the additional transformation $p_\euc \rightarrow -p_\euc$~\cite{Roberts:1994dr}, up to an overall sign that may be reabsorbed in the definition of the propagator. 
With this extra step, we obtain
\begin{align}
    \left[\overline{\fatg}_\mu(q,p_2,-p_1)\gamma^\mu\right]_\wic = (d-1)\left[ \lambda^\euc_1(q^\euc,p^\euc_2,-p^\euc_1)+i \lambda^\euc_3(q^\euc,p^\euc_2,-p^\euc_1)(\slashed{p}^\euc_1+\slashed{p}^\euc_2) \right]+ \ldots\,.
\end{align}
The above result is exactly what we obtain when contracting the 
Euclidean form in \1eq{decomp2} with $\gamma^\mu_\euc$ from the right, establishing that \2eqs{decomp2}{mittbas} is the Euclidean equivalent of \2eqs{decomp}{Taus}.

The same procedure can be applied in the soft-gluon configuration, to relate our form factors $\lambda_i^\sg(p^2)$ with the $\lambda_j^{\rm L}(p_\euc^2)$ computed on the lattice study of~\cite{Kizilersu:2021jen}.

Specifically, in the soft-gluon limit, \1eq{decomp} reduces to
\begin{align} 
\label{softgluon1}
    \overline{\fatg}_\mu(0,p,-p)= \gamma_\mu\lambda_1^\sg(p^2) + 2p_\mu\lambda_2^\sg(p^2) +4 \slashed{p}\,p_\mu \lambda_6^\sg(p^2)\,.
\end{align}
Note that, although in this kinematic limit the tensor $\tau_{3}^\nu(p,-p) = 2\slashed{p}\gamma^\nu$ is non-zero, the associated form factor $\lambda_3^\sg(p^2)$ vanishes due to charge conjugation symmetry, as discussed below \1eq{conjC}.

Then, contracting \1eq{softgluon1} with $\gamma^\nu$, passing to Euclidean space through \1eq{euc_transf}, and using $p_\euc \rightarrow -p_\euc$ to account for the convention $S^{-1}_{0\euc}(p_\euc) = i \slashed{p}_\euc+m_q$ used in~\cite{Kizilersu:2021jen}, yields
\begin{align} \label{gamma2}
    \left[\overline{\fatg}_\mu(0,p_\euc,-p_\euc)\gamma^\mu\right]_\wic= d\lambda_{1\rm{\s{E}}}^\sg(p_\euc^2)  -2i\slashed{p}_\euc\lambda_{2\rm{\s{E}}}^\sg(p_\euc^2) -4 p_\euc^2 \lambda_{6\rm{\s{E}}}^\sg(p_\euc^2)\,,
\end{align}
where we note that the last equality in \1eq{euc_transf} implies $\lambda_{i\rm{\s{E}}}^\sg(p_\euc^2)=\lambda_i^\sg(-p_\euc^2)$.

Finally, substituting \1eq{gamma2} into \1eq{Gammag}, it follows that the Minkowski basis for the soft-gluon kinematics in \1eq{softgluon1} is equivalent to decomposing the vertex directly in Euclidean space as
\begin{align} 
\label{softgluoneuc}
    \overline{\fatg}^\euc_\mu(0,p_\euc,-p_\euc)= \gamma^\euc_\mu\lambda_{1\rm{\s{E}}}^\sg(p_\euc^2) - 2ip^\euc_\mu\lambda_{2\rm{\s{E}}}^\sg(p_\euc^2) -4 \slashed{p}^\euc p^\euc_\mu \lambda_{6\rm{\s{E}}}^\sg(p_\euc^2)\,.
\end{align}
The above is precisely the same as the basis used in the lattice study of~\cite{Kizilersu:2021jen}, with the identification
\begin{align}   
\label{lattice1}
\lambda_{1{\s{\rm{E}}}}^\sg(p_\euc^2)=\lambda_1^{\rm L}(p_\euc^2) \,, \quad \lambda_{6{\s{\rm{E}}}}^\sg(p_\euc^2)=\lambda^{\rm L}_2(p_\euc^2)  \,, \quad \lambda_{2{\s{\rm{E}}}}^\sg(p_\euc^2)=\lambda^{\rm L}_3(p_\euc^2)\,.
\end{align}



%

\end{document}